# TEC evidence for near-equatorial energy deposition by 30-keV electrons in the topside ionosphere

A. V. Suvorova[1,3], A. V. Dmitriev[2,3], L.-C. Tsai[1], V. E. Kunitsyn[4], E. S. Andreeva[4], I. A. Nesterov[4], and L. L. Lazutin[3]

[1]Center for Space and Remote Sensing Research, National Central University, Jhongli, Taiwan

[2]Institute of Space Science, National Central University, Jhongli, Taiwan

[3]Skobeltsyn Institute of Nuclear Physics, Lomonosov Moscow State University, Moscow, Russia

[4] Faculty of Physics, Lomonosov Moscow State University, Moscow, Russia

Short title: ELECTRON IMPACT ON THE IONOSPHERE

**Abstract**

Observations of energetic electrons (10 - 300 keV) by NOAA/POES and DMSP satellites at heights <1000 km during the period from 1999 to 2010 allowed finding abnormal intense fluxes of ~$10^6$ - $10^7$ $cm^{-2}$ $s^{-1}$ $sr^{-1}$ for quasi-trapped electrons appearing within the forbidden zone of low latitudes over the African, Indo-China, and Pacific regions. Extreme fluxes appeared often in the early morning and persisted for several hours during the maximum and recovery phase of geomagnetic storms. We analyzed nine storm-time events when extreme electron fluxes first appeared in the Eastern Hemisphere, then drifted further eastward toward the South-Atlantic Anomaly. Using the electron spectra, we estimated the possible ionization effect produced by quasi-trapped electrons in the topside ionosphere. The estimated ionization was found to be large enough to satisfy observed storm-time increases in the ionospheric total electron content determined for the same spatial and temporal ranges from global ionospheric maps. Additionally, extreme fluxes of quasi-trapped electrons were accompanied by the significant elevation of the low-latitude F-layer obtained from COSMIC/FORMOSAT-3 radio occultation measurements. We suggest that the storm-time *ExB* drift of energetic electrons from the inner radiation belt is an important driver of positive ionospheric storms within low-latitude and equatorial regions.

# 1. Introduction

The coupling between the magnetosphere and the ionosphere through precipitating particles from radiation belts and the central plasma sheet that strongly controls ionization and conductance in the ionosphere has been known for some time, although it is one of the most difficult couplings to understand [*e.g. Paulikas*, 1975; *Buonsatto*, 1999; *Ma et al.*, 2008; *Kelley*, 2009]. The energy deposition of precipitating particles of various energies mainly through the ionization of atmospheric atoms occurs at different altitudes, particularly within the D, E, and F regions of the ionosphere and extends from high to mid-latitudes and over the South-Atlantic Anomaly (SAA) at low latitudes [e.g., *Voss and Smith*, 1980; *Vampola and Gorney*, 1983; *Rees et al.*, 1988; *Abel et al.*, 1997]. An impact of precipitating particles in the lower ionosphere (the D and E layers) and the upper atmosphere in terms of enhanced ionization and other related aeronomical effects has been investigated in a large number of studies [e.g., *Buonsatto*, 1999 and reference therein; *Nishino et al.*, 2002; *Peter et al.*, 2006; *Clilverd et al.*, 2008; 2010; *Turunen et al.*, 2009; *Lam et al.*, 2010; *Dmitriev et al.*, 2011], while evidence for ionospheric signatures in the topside ionosphere (F region) in the mid-latitudes and over the SAA at low latitudes have also been reported [e.g., *Foster et al.*, 1994; 1998; *Abdu et al.*, 2005; *Kunitsyn et al.*, 2008a;b; *Dmitriev and Yeh*, 2008; *Pedatella et al.*, 2009; *Ngwira et al.*, 2012].

Since the beginning of the space era, the fluxes of energetic electrons with energies of a few tens of keV have been observed in the ionosphere in numerous experiments (see review of *Paulikas* [1975]). Electron fluxes have been found to increase with altitude and with geomagnetic activity [*Hill et al.*, 1970; *Goldberg et al.*, 1974]. Three populations of electrons, trapped, precipitating, and quasi-trapped have been determined [*Konho*, 1973]. The classification is based on the physical behaviour of electrons with different local equatorial pitch angles (the angle between the velocity of a particle and the magnetic field line) [*e.g., Tu et al.*, 2010; *Rodger et al.*, 2010]. Notably, the "precipitating" or "un-trapped" particles have local equatorial pitch angles that range within a bounce loss cone. Due to scattering in the atmosphere, particles are lost within one bounce period since their mirror points are below an $H$min of ~100 km. Particles that close their drift path around the Earth are referred to as stably "trapped". Particles from a "quasi-trapped" (or locally trapped) population cannot close the full drift shell around the Earth and the pitch angles of quasi-trapped particles range within the drift loss cone. Such particles can make a number of bounces, however, at a certain longitude, their local equatorial pitch angle occurs within the bounce loss cone and, as a result, the particles are precipitated.

The energy spectra of storm-time magnetospheric particles, from which the energy flux is derived, are well-known for the population in the radiation belt and the auroral zone (i.e. mainly for L shells >2). Under geomagnetically disturbed conditions, the flux of trapped and precipitating particles increases by several orders of magnitude. Established is that the main contributor to the redundant ionization of the lower ionosphere within the auroral zone (>60° latitude, L>7) is precipitating electrons with energy from 1 to 100 keV, while the energy contribution from soft electrons <1 keV is relatively small [*Ostgaard et al.*, 2001]. During geomagnetically active periods, 1 - 30 keV electron precipitations increase by three orders of magnitude (i.e. approximately above $10^6$ units, where unit means electrons/(cm$^2$ s sr)). Therefore, the energy flux deposited into the atmosphere is expected to increase from 1 to 30 mW/m$^2$ [e.g., *Chenette et al.*, 1993; *Rees et al.*, 1988]. *Ostgaard et al.* (2001) found that energy deposition does not occur uniformly in different regions of space. In the evening sector, it is primarily due to electrons below 10 keV, while in the pre-midnight and morning sectors, it is provided by energetic electrons of higher energies.

Widely accepted is that the main source of energetic electrons in the inner magnetosphere at ionospheric altitudes from ~100 to ~2000 km is the inner and outer radiation belts (the IRB and ORB, respectively). The precipitating population of the RB reaches the ionosphere and the atmosphere at mid-latitudes from 20° to 60°, while trapped particles reach ionospheric altitudes only in the vicinity of the SAA at lower latitudes. The energy of electrons varies from a few tens of keV to several MeV. The bottom of the IRB at the equator is located at an L-shell of ~1.2 and its height changes from 1,600 to ~ 300 km depending on longitude. Detailed characteristics for electron precipitations from the inner magnetosphere (L~2 - 7) during storms are now available over a wide energy range from 100 eV to MeV. Precipitation from the ORB is slightly less intense than that within the auroral zone and less than the trapped flux, while precipitation from the IRB, in-turn, is less than that from the ORB and exhibits less dramatic variability during storms.

Essential information for the evaluation of energy deposition into the upper atmosphere and ionosphere is provided by recent statistical results on the suprathermal- (0.1 - 10 keV), the medium- (>30 keV), and high- energy (>300 keV) electron distributions in the ORB [*Li et al.*, 2010; *Lam et al.*, 2010; *Rodger et al.*, 2010; *Clilverd et al.*, 2010; *Lazutin*, 2012], and high-energy electrons in the IRB [*Tadokoro et al.*, 2007; 2009]. *Li et al.* [2010] found that during the geomagnetically active period, the energy flux of suprathermal electrons increased inside of the plasmapause (L<4), particularly from the pre-midnight to the dawn sectors, and that 10-keV electrons formed a stable electron ring distribution. *Lam et al.* [2010] reported that >30 keV precipitating electron fluxes are maximized during substorm activity outside of the plasmapause on the dawnside, and exceed 5 x $10^5$ units that extend over a larger range of local time from pre-midnight dawn until noon. *Lazutin* [2012] investigated less known phenomena such as the dawn-dusk asymmetry of high-energy electrons during storms, where an enhanced particle flux for the ORB is detected in the dawn sector. Studies by *Tadokoro et al.* [2007; 2009] of high-energy electrons in the IRB (2 < L < 2.5) indicated that the IRB is not only dramatically disturbed during a major storm, but that it is quite variable in response even to moderate magnetic activity (Dst > -100 nT), contrary to the general view. During moderate magnetic storms researchers have determined that precipitating high-energy electrons exhibit sudden enhancements of over one order of magnitude.

Focusing our work on the particle effect in the topside ionosphere (the F region), we list studies that provide important information regarding the F-region signatures that appeared due to the direct ionization of thermospheric species by energetic electrons [*Foster et al.*, 1998; *Pedatella et al.*, 2009; *Dmitriev and Yeh*, 2008; *Kunitsyn et al.*, 2008a,b; *Ngwira et al.*, 2012]. Such studies demonstrate features such as the uplift of the F region, and spots of enhanced ionization in the bottom-side F-layer at mid-latitudes and in the topside region over the SAA. The contribution of the particle effect indicates that an increase in the Total Electron Content (TEC) at the mid-latitudes can be approximately 10 TECU (1TECU = $10^{12}$ electrons/cm$^2$) as inferred, for example, from the study by *Pedatella et al.* [2009].

A quasi-trapped population of energetic electrons was observed below the IRB at an L < 1.2 within the near-equatorial region. While its existence has been known for a long time [*e.g., Krasovskii et al.*,1958; 1961; *Galperin et al.*, 1970; *Hill et al.*, 1970; *Heikkilla*, 1971; *Hayakawa et al.*, 1973; *Kohno*, 1973], information for fluxes and spectra has been scarce and controversial. In general, the flux of energetic electrons at L < 1.2 is invariably weak and certainly less than those inside of the IRB, ORB, or auroral zone during quiet as well as storm times. Therefore, it is widely accepted that the particle impact at low latitudes is insufficient to produce appreciable ionization [e.g., *Paulikas*, 1975; *Voss and Smith*, 1980; *Kelley*, 2009]. On the other hand, from the second spaceship experiment, *Savenko et al.* [1962] found sporadic events with unusually large fluxes of ~10 keV electrons within the equatorial ionospheric F-region at ~320 km outside of the SAA, in the region between 150°E and 150°W longitude. Electron fluxes were comparable to those in the radiation belts. Since these electrons sink into the SAA on their eastward drift path at low altitudes and L-shells, a full drift period was not possible. Therefore, the electrons are regarded as quasi-trapped. Later, *Leiu et al.* [1988] found flux increases and the appearance of an energy spectrum peak at ~10 keV for precipitating low-energy electrons during the disturbed period. The spectrum changed from a typical "power law" to an "outburst". The flux increase was approximately ~100 times greater than the quiet level and 10 times greater than the flux within the SAA. Such features were observed at an ~ 240 km height over the Indochina and Pacific regions and located below the edge of the IRB at an L ~

1.16 (i.e. coinciding with the areas revealed by *Savenko et al.* [1962]).

Modern experiments measure energetic electron fluxes below and inside the IRB. During quiet or weakly disturbed geomagnetic activity, the medium-energy population of the IRB has complex spatial and energy band structures [e.g., *Datlowe et al.*, 1985; *Kudela et al.*, 1992; *Biryukov et al.*, 1998; *Grachev et al.*, 2005; *Sauvaud et al.*, 2006; 2013; *Grigoryan et al.*, 2008]. *Evans* [1988] first reported dramatic enhancements of quasi-trapped energetic electron fluxes at an ~850 km in response to large magnetic storms; thereafter, other studies also determined intense fluxes at low-altitudes near the equator [*Tanaka et al.*, 1990; *Pinto et al.*, 1992; *Gusev et al.*, 1995; *Asikainen and Mursula*, 2005; *Suvorova et al.*, 2012]. The observed phenomena favour continued confidence in past observations within lower energy ranges and for lower altitudes [*Krasovskii et al.*, 1961; *Savenko et al.*, 1962; *Knudsen*, 1968; *Heikkila*, 1971; *Goldberg et al.*, 1974; *Leiu et al.*, 1988]. However, the phenomena have not been completely investigated and we still have no explanation for the appearance of quasi-trapped electrons at such low L-values (< 1.2).

*Evans* [1988] suggested that the observed quasi-trapped >30 keV electrons were injected westward to the SAA at night then drifted eastward toward dawn. He also noted a short lifetime for electron enhancements of roughly a few hours, indicating the transient nature of electron injections. According to *Asikainen and Mursula* [2005], energetic electrons were injected into the pre-midnight sector to L ~ 1.14 at the end of the main phase of a major magnetic storm. Maximum fluxes were only seen inside of the SAA, suggesting the charge exchange of ring current ions as a possible mechanism for electron injection.

The above mentioned observations indicated that increases in quasi-trapped electrons with energies of a few tens of keV were comparable with auroral zone intensities of >$10^6$ units. Therefore, it is reasonable to expect their significant ionization impact in the ionosphere.

Recently, a comparative analysis of quasi-trapped electron phenomena with a positive phase of ionospheric storms, as observed by the COSMIC/FS3 satellite, was made by *Suvorova et al.* [2012] indicating a strong relationship between the two phenomena and suggesting energetic particles as a storm-related ionization source for the topside ionosphere over a wide range of longitudes in the Taiwan – Japan region through the Pacific to the SAA. The energy flux of electrons was determined to be ~4.5 mW/m$^2$, allowing abundant ionization of 20 TECU at low and mid-latitudes.

The purpose of this work is to provide a numerical evaluation of the effect of energetic electrons in the ionosphere. Section 2 describes the method and the uncertainties surrounding estimations of ionospheric ionization. Section 3 provides a data source description. Section 4 is devoted to a statistical analysis of the energetic electron enhancements observed by fleets of NOAA/POES and DMSP satellites at altitudes of ~850 km (i.e. L < 1.15) over the time period from 1999 to 2006. In Section 5, we consider case events and demonstrate that under certain electron injections into the inner magnetosphere, positive ionospheric storms develop. The results are discussed in Section 6. Section 7 provides conclusions.

## 2. Quasi-trapped electron impacts and uncertainties in TEC estimations

As discussed above, during major storms at heights lower than the IRB edge (±20°, L <1.2), quasi-trapped energetic electrons exhibit a sudden enhancement of several hours, and are very large against a simultaneous increase in precipitating electrons for which the flux is smaller by three orders of magnitude. Considerable enhancements in >30 keV quasi-trapped electrons near the equator are comparable to the electron intensity within the SAA region and in the auroral zone, where the importance of the ionizing particle effect in the ionosphere is well established. However, important information regarding quasi-trapped electrons such as the energy spectra, the pitch-angle, and the spatial and local time distributions are still incomplete. Therefore, it is particularly difficult to properly estimate the possible ionizing effect produced by energetic electrons.

### 2.1. Grounds

We estimated the spatial region where electrons lose their energy in the ionization of the atmospheric atoms. A precipitating electron with an energy of ~30 keV and a zero pitch angle is able to reach altitudes of ~90 km [e.g., *Dmitriev et al.*, 2010]. However, using NOAA/POES observations we determined that the vast majority of enhanced electrons are quasi-trapped with pitch angles close to 90° [*Suvorova et al.*, 2012]. Such electrons bounce along magnetic field lines about the top points of the magnetic field lines. The bouncing motion is very fast and has a period of a portion of a second.

Due to the asymmetrical orientation of the geomagnetic dipole, the height of the top points varies with longitude. Figure 1 shows the longitudinal variation of the height of the drift of L-shells calculated for the geomagnetic equator using the IGRF model for the 2005 epoch. Participating in a gradient drift, energetic electrons move eastward along the drift shells. L-shells descend beginning with the region of Indochina at longitudes of ~120°. In this region, the height of ~900 km corresponds to an L ~ 1.05. Above the Pacific region, the altitude of the top points for bouncing electrons decreases with increasing longitude. The drift shells rich minimal heights in the SAA region, where practically all of the particles, trapped at L < 1.1, are lost. At altitudes of 1,000 km and below, the period of azimuthal drift for 30 keV electrons is ~20 hours [*Lyons and Williams*, 1984]. Hence, electrons with large pitch angles can make many thousands of bounces before they are lost within the SAA region.

Taking into account the specific ionization of electrons (the energy loss per unit distance) and the standard vertical profile of the upper atmosphere [e.g., *Dmitriev et al.*, 2008], we were able calculate the number of bounces between the top point at 800 km and the mirror points at a height of $H$min until the ~30 keV electron lost its energy in ionization. In this manner, we found that for a $H$min below 600 km, ~30 keV electrons lose their energy within ~2 hours. During this time period, electrons pass eastward no more than ~30° in longitude because of a very slow azimuthal drift. Hence, quasi-trapped electrons, as observed westward from a longitude of 150°W, had a large chance of losing their energy during ionization within the ionosphere. Due to an arched magnetic field configuration, this ionization is released in a region of geomagnetic latitudes above ~20°. Important to note is that charged particles spend most of their time in the vicinity of the mirror point. Hence, quasi-trapped electrons lose most of their energy at low to middle latitudes rather than at the equator, corresponding well with the spatial location of low/mid-latitude positive ionospheric storms.

*2.2 Uncertainties*

Prior to calculating the ionizing electron impact we needed to consider factors that influenced TEC estimation uncertainties.

Actual distributions of electron pitch-angles and energy losses are unknown in processes such as ionization, excitation, and secondary electron production. In the topside ionosphere, the variability range of the electron recombination rate is still not well established. In this region, the averaged electron content and its variation are also purely known.

**Pitch-angle distribution.** From our study of the energetic electron flux distribution in the near-equatorial region during major storms [*Suvorova et al.*, 2012], we know that particles locally trapped at a given longitude (with a pitch angle of ≤90°) exhibit large increases, while particles with a pitch angle within the local bounce loss cone, it can be said, do not. Therefore, we assumed an anisotropic pitch-angle distribution for the electrons and arbitrarily used a multiplying factor of 2π instead of the 4π valid for isotropic distribution.

**Energy losses during ionization.** In addition to ionization, a part of the primary energy of energetic electrons is lost for the excitation of thermospheric species and for secondary electron production. Moreover, these processes can alter the ionospheric composition during energetic electron impact and can, in turn, influence the chemical reaction rate, and, in particular, the recombination rate [*Sheehan and St.-Maurice*, 2004]. Since these complex processes remain incompletely studied and since the distribution of electron energy losses in the ionosphere are currently difficult to determine, information regarding electron losses inside the topside ionosphere are not available [*e. g., Sheehan and St.-Maurice*, 2004]. Hence, we arbitrarily assumed that electrons lose 100% of their energy during the ionization of oxygen atoms, which dominate the topside ionosphere. Using an oxygen atom for first ionization potential of 13.6 eV, we obtained the upper limit value for the ion-electron pairs produced by energetic electrons.

Also, at heights from 300 to 600 km, the total thermospheric density decreased by one order of magnitude, by ~200 km [*Schunk and Nagy*, 1980], indicating a reducing efficiency for ionization with height. On the other hand, this effect should be offset with the surplus of time spent by quasi-trapped particles at these altitudes (see above). In total, we suggest that the calculated TEC may need to be decreased as a result of this factor, likely no more than roughly several percent.

***Recombination rate.*** The ionization balance in the ionosphere is largely controlled by production (e.g., solar EUV flux), transport (e.g. thermospheric winds), and loss (e.g., recombination) processes (see *Schunk and Nagy* [1980] for a review). For the current study, we needed to know a decay rate for ionospheric electron density as predicted by atmospheric ion-molecule chemistry, in particular, for the height interval from 300 to 600 km within the topside F region. Note that above 600 km, the concentration of atmospheric species, especially molecular ions, is so small that relevant aeronomic processes can be neglected. However, at heights below 600 km, the elevated content of molecules and molecular ions decreases the electron density within the F region, as clearly shown for the high-latitude region associated with electron precipitation in the auroral zone and for high-, mid-, and low-latitude ionospheric troughs (see *Campbell et al.* [2006] for references). Important to emphasize here is that basic recombination and excitation rates strongly depend on thermospheric species, temperature, and electron density [e.g., *Danilov and Ivanov-Kholodnyi*, 1965]. In the auroral zone, for example, temperatures depend on the spectrum of precipitating electrons [*e.g. Vlasov and Kelly*, 2003]. Due to the large temperature variability range (a few times) during thermospheric perturbation, one can obtain a recombination rate uncertainty with a factor of approximately two. Additionally, as a result of large differences in the approximation for the basic reaction rate coefficients found for different studies that result in a recombination rate uncertainty of as much as a factor of two, the uncertainty increases (see details in *Sheehan and St.-Maurice* [2004]). Therefore, we were faced with the problem that the decay rate can differ by an order of magnitude. On the other hand, an investigation of the chemical composition effect in the ionosphere revealed a significant increase (of a few times) for the recombination coefficient during thermospheric perturbation. For example, from modeling the low-latitude ionosphere under composition changes, *Jenkins et al.* [1997] found a reduction in the electron density of two thirds under an undisturbed chemical composition and noted that higher electron temperatures and densities lead to a lower total ionization. A study of the high-latitude trough [*Vlasov and Kelly*, 2003] also indicated that during intensive electron precipitation, when chemical compositions and thermospheric species temperatures strongly change, the recombination rate at a height of 300 km can increase at least four times as compared to a normal value of $2 \times 10^{-4}$ s$^{-1}$ (note that a comparison without any reserve indicates an increase up to 10 times). *Mishin et al.* [2004] used values of $>1.7 \times 10^{-3}$ s$^{-1}$ and noted that as the F peak density decay rate of $>1.7 \times 10^{-3}$ s$^{-1}$ increased with electron density, ''fresh'' plasma increased the ''processing'' rate upon arrival. Additionally, it was noted by *Vlasov and Kelly* [2003] that due to their penetration to lower altitudes and the subsequent production of secondary electrons with energies of 2–3 eV, electrons with energies of ~30 keV induce an enhancement in nitric oxide. Such electrons excite $N_2$ vibrational levels, strongly increasing NO production, leading to an increase in the recombination rate. Therefore, since energetic electrons impacting the topside ionosphere cause the recombination rate to increase, we used a recombination coefficient of $5 \times 10^{-3}$ s$^{-1}$ in our calculations.

### 2.3. The problem of quiet days

Geomagnetic activity practically never ceases, resulting in the day-to-day variability of the TEC. In general, the averaged pattern for the five quietest days in a month or even on a previous day before storm onset are often used in the large majority of ionospheric studies for calculating storm-related TEC changes. We believe that a consistent choice for the quiet-time period is one of the most important factors. As a result, we applied an enhanced set of criteria for determining the quiet state for the magnetosphere-ionosphere system. Initially, we selected a few quiet, 24-hour intervals surrounding a storm day on the basis of the geomagnetic indices of AE <100 nT and SYM-H > -20 nT. We also excluded days with a high solar activity manifested by X-class flares and SEP events, as monitored by GOES satellites. As a rule, we designated a few quiet days for each storm and could use any one or all (averaged) of the values for obtaining a reference quiet-day pattern. However, a comparative analysis of quiet states in the ionosphere-magnetosphere system indicated that differences between various "quiet" radiation belts and "quiet" ionospheric TEC patterns were very noticeable. For example, the ionospheric background level varied within the range from 5 - 10 TECU. Such uncertainty in the quiet level became crucial for searching the storm-time particle effect, estimated to be ~20 TECU [*Suvorova et al.*, 2012].

To minimize uncertainty, we used additional criteria for the appropriate solar wind parameters, supporting "the quietest ionosphere-magnetosphere system" [*Tsurutani et al.*, 2006]. Additional constraints were imposed on the IMF fast variation, the high solar wind velocity, and the sharp increase in solar wind dynamic pressure that can cause weak auroral activity, strong magnetospheric compression, and that can also disturb the radiation belt (increases in particle fluxes). Actually, one or two days during each ~27-day interval were characterized by the lowest solar wind speed that corresponds to the quietest state of the ionosphere-magnetosphere system [*Tsurutani et al.*, 2006]. Hence, this method of quiet day selection allowed a viewing of the ionization effect (if any) for energetic electrons with minimal enhancements of approximately ~5 TECU.

## 3. Data Sources and Methods

### 3.1. Particle data

We used the time profiles for >30 keV electrons fluxes measured by the polar orbiting NOAA/POES satellite fleet [*Rodger et al.*, 2010]. POES satellites have Sun-synchronous orbits at altitudes of ~ 800 - 850 km (with ~100 minute periods of revolution). The first POES satellite, NOAA-15, began operation and supplied radiation belt data from 1998. POES data are available at http://poes.ngdc.noaa.gov/data/. The orbital planes of the POES satellites NOAA-15, NOAA-16, NOAA-17, NOAA-18, and METOP-02 (hereafter, P5, P6, P7, P8, and P2, respectively) from 2001-2006 are approximately 0700 to 1900 LT, 0200 to 1400 LT, 1000 to 2200 LT, 0200 to 1400 LT, and 0930 to 2130 LT, respectively.

The Medium Energy Proton and Electron Detector (MEPED) onboard POES satellites includes two identical electron solid-state detector telescopes and measures particle fluxes in two directions - along and perpendicular to the local vertical direction. For a polar-orbiting satellite, the 0º-telescope is pointed almost to the Zenith and the 90º-telescope is oriented in the north-south direction [*Huston and Pfitzer*, 1998; *Evans and Greer*, 2004]. Hence, at low latitudes the 0º-telescope mainly measures quasi-trapped particles and the 90º-telescope measures precipitating particles, and vice versa at high latitudes. Hereafter, as a default, we use the terms "quasi-trapped" and "precipitating" with respect to equatorial latitudes.

Experimental data for electrons in the low-energy range from 30 eV to 30 keV, as measured by the SSJ/4 particle detectors onboard the Sun-synchronous polar orbiting DMSP fleet, have also been used to substantiate POES observations. DMSP particle spectrograms are provided online by the Auroral Particle and Imagery Group of JHU/APL (http://ccmc.gsfc.nasa.gov/models/modelinfo.php?model=AACGM&type=1). The altitude of DMSP satellites is 840 km.

The radial profiles of the electron belt were analyzed using data from the polar orbiting satellite SERVIS-1 [*Kodaira et al.*, 2005], launched in December 2003 on a Sun-synchronous (inclination 99.5°) orbit in the dawn-dusk plane at an the altitude of 1000 km. The data are considered reliable until March 2005. Electrons in the energy range from 0.3 - 1.5 and 1.7 - 3.4 MeV were detected by a telescope with a 60° field of view.

### 3.2. Ionospheric data

**HORT and LORT imaging using beacon data.** The first experiments on ionospheric tomography utilized the 150-MHz and 400-MHz radio signals of low-orbiting (LO) navigation satellite systems such as the American «Transit» - Naval Navigation Satellite System (NNSS) or the Russian "Tsykada" and "Parus". The application of these satellite systems to ionospheric research was first suggested by *Austen et al.* [1988] for imaging the 2D distribution of electron density. The first experimental tomographic reconstructions (2D ionospheric cross sections) were obtained in 1990 [*Andreeva et al.*, 1990]. Over the past few decades, more than ten tomographic systems have been constructed, and have provided extensive new information regarding the structure of the ionosphere. Findings are described in a number of reviews and books [*Kunitsyn and Tereshchenko*, 1992; *Leitinger*, 1999; *Kunitsyn and Tereshchenko*, 2003; *Pryse*, 2003; *Kersley*, 2005; *Bust and Mitchell*, 2008].

In June 1994 the National Central University of Taiwan deployed the Low-latitude Ionospheric Tomography Network (LITN) that included six ground stations for receiving NNSS signals. LITN stations span a latitudinal interval of 16.7° (from 14.6°N to 31.3°N) within a longitudinal band from 1° to 121°E. LITN ionospheric tomography networks extend along the Taiwan meridian and include receiving stations with data acquisition systems that were developed by the National Central University and the University of Illinois at Urbana-Champaign in 1994. The six receiving stations are, from south to north, Manila (121.0°E, 14.6°N), Baguio (120.5°E, 16.4°N), Kaohsiung (121.0°E, 22.5°N), Chungli (120.6°E, 25.0°N), Wenzhou (120.6°E, 28.0°N), and Shanghai (121.5°E, 31.3°N). Therefore, the LITN was specifically designed to observe large-scale ionospheric structures over the northern equatorial region using a tomographic imaging technique applied to NNSS satellite beacon data [*Huang et al.*, 1999; *Andreeva et al.*, 2000; *Yeh et al.*, 2001; *Franke et al.*, 2003]. As a result of further operation, the LITN has functioned with a different number of receivers. During some intervals (e.g., during 2006), measurements were only performed with the receiver in Taiwan [*Kunitsyn et al.*, 2008a and b].

High orbital (HO) radio tomography (RT) (HORT), which utilizes radio transmissions from high orbital GPS and GLONASS navigational systems measured from a ground receiving network, is another method of ionospheric radio tomography. All of these HO satellite systems are referred to as GNSS (Global Navigation Satellite Systems). At present, information is continuously provided by GNSS receiving networks and is used for reconstructing the distributions of electron density in the ionosphere. Several regional and global networks of GNSS receivers exist, including the IGS (International Geodetic Service), which includes more than 2,000 receivers.

HORT enables 3D or 4D imaging (a 3D distribution every hour or half an hour) for electron density. A specific feature of the inverse problems of radio sounding based on GNSS data, relating to tomographic problems with incomplete data, is a large dimensionality. Since the angular velocity of high orbiting GNSS satellites is relatively low, an allowance should be made for time variations of the ionosphere leading to the four-dimensional statement for RT problems (three spatial coordinates and time). Due to the four-dimensionality of the problem, incompleteness of the data becomes the most essential factor - not all of the points in space are covered by satellite-to-receiver rays, producing data gaps in regions with a small number of receivers. The solution to this problem requires a special approach [*Kunitsyn et al.*, 2005, 2010, 2011; *Nesterov and Kunitsyn*, 2011]. The spatial resolution of HORT is significantly lower than that of LORT - typically, the vertical and horizontal resolution is roughly 100 km.

***Global ionospheric maps.*** Global ionospheric maps (GIM) of vertical total electron content (VTEC) were obtained from a world-wide network of ~200 ground based receivers [*Rebischung et al.*, 2012; http://aiuws.unibe.ch/ionosphere/]. GIMs were provided by the International Global Navigation Satellite Systems Service (IGS) and other institutions [http://igscb.jpl.nasa.gov/network/refframe.html]. The set of ground-based stations was selected by the IGS Reference Frame Working Group. The main selection criteria were station performance, track record, monumentation, collocation, and geographical distribution. The latter refers to the uniform distribution of stations around the globe. VTEC was modeled in a solar-geomagnetic reference frame using a spherical harmonic expansion up to a degree and an order of 15. In the time domain, piece-wise linear functions were used for representation. The statistical error value of VTEC provided by CODE is 0.1 TECU. The time spacing of their vertices is 2 hours, conforming to the epochs of VTEC maps [ftp://ftp.unibe.ch/aiub/CODE/].

*COSMIC/FORMOSAT-3 radio occultation tomography.* Measurements of COSMIC/FORMOSAT-3 space-borne experiments were used for constructing vertical profiles of electron content (EC). Six satellites of the COSMIC/FS3 mission produced a sounding of the ionosphere on the basis of the radio occultation (RO) technique, making use of radio signals transmitted by GPS satellites [*Hajj et al.*, 2000]. A 3D EC distribution was deduced through relaxation using red-black smoothing on numerous EC height profiles. Such 3D EC imaging was used as an initial guess for beginning the iterative Multiplicative Algebraic Reconstruction Technique (MART) algorithm. The 3D tomography of the EC was then produced for the entire globe with a time step of 2 hours and a spatial grid of 5º in longitude, 1º in latitude, and 5 km in height [*Tsai et al.*, 2006].

## 4. Storm-time quasi-trapped electrons: statistics

We analyzed the quasi-trapped >30 keV electron fluxes measured onboard NOAA/POES satellites near the equator during major geomagnetic storms ($Dst \leq -100$ nT) occurred from 1999 to 2006. From more than 60 storms we only selected nine storm-events for which integral directional electron fluxes $F(>E)$ exceeded $10^6$ units near the equator at any longitude and above the entire Eastern Hemisphere. We did not include events with flux enhancements that were observed only in the Western Hemisphere. Table 1 specifies the days for major storms when extremely intense fluxes appeared at the eastern longitudes and for the quiet days that were selected using the method described in Section 2. Interesting is that the salient feature, "eastern longitude region", was not seen during many of the super-storms ($Dst < -200$ nT), for example, 7 November 2004, 20 November 2003, or 11 April and 31 March 2001.

Figure 2 shows four summary patterns for the >30-keV electron flux intensity obtained by accumulating data over multiple orbits of satellites for quasi-trapped/precipitating electrons during storm and quiet periods. Quiet patterns were characterized by moderate fluxes of trapped electrons with more intense fluxes inside the SAA, higher than $10^4$ units, weak electron precipitation everywhere outside the SAA, and a negligible intensity for quasi-trapped electrons with a value below $10^2$ units. In contrast, during the disturbed period, all of the electron populations were significantly intensified, including those that were quasi-trapped.

The two patterns on the left-side of Figure 2 demonstrate the well-known general characteristics of dramatic changes of the magnetospheric state during geomagnetic storms (i.e. an enhancement of both trapped and precipitating electron fluxes at high/mid-latitudes and within the SAA area). For this case, the equatorward boundary of the auroral oval and the outer radiation belt moved toward the sub-auroral and mid-latitudes, respectively, with an extreme increase in electron intensities. A more interesting feature was revealed outside of the SAA area near the equator during these major storms. Unusually large enhancements of the quasi-trapped electron flux were determined within the forbidden range of drift shells (top-left). Very intense fluxes of quasi-trapped electrons over Indochina and the Pacific Ocean were comparable with the auroral fluxes. Meanwhile, the equatorial enhancement of precipitating electrons was weaker by 2-4 orders of magnitude (bottom-left). From Table 1 (the last column) one can see that the appearance of great flux enhancements within the Eastern Hemisphere was observed during various storm phases. We found that in some cases, particularly during the storm recovery period, a sharp solar wind dynamic pressure increase can cause a strong magnetospheric compression, resulting in quasi-trapped electron enhancements.

Furthermore, we estimated values for energy deposition for quasi-trapped energetic electrons in relation to each storm-time event. For this purpose, we chose a specific time interval, when >30 keV quasi-trapped electron fluxes were maximized at some eastern longitude then determined directional energy flux spectra using mid- and high-energy electron measurements (>30; >100; >300 keV). Integral and differential spectra were obtained by fitting integral electron fluxes $F(>E)$ using a power law, $F(>E, \text{keV}) = J_0 \cdot E^{-\alpha}$, or an exponential law, $F(>E, \text{keV}) = J_0 \cdot \exp(-E/E_0)$. The integrated energy flux density, $JE$ (eV/cm$^2$ s sr) was calculated using the following expressions:

$$JE = 10^3 \cdot \alpha/(\alpha - 1) \cdot J_0 \cdot E^{-\alpha+1}$$
$$JE = 10^3 \cdot (E + E_0) \cdot J_0 \cdot \exp(-E/E_0)$$
(1),

where $\alpha$, $E_0$, and $J_0$ are the fitting parameters for the integral spectrum.

The resultant value for the $JE$ was obtained for the energy range from 30-100 keV. We also used complementary information regarding the energy flux density, $JE$, provided online for the DMSP particle spectrogram of low-energy electrons in the range from 30 eV to 30 keV. By multiplying the $JE$ value by a factor of $2\pi$ (see Section 2) we obtained an arbitrary maximal value for the energy flux, $IE$ (erg/cm$^2$ s = mW/m$^2$). Tables 2 and 3 provide the resultant values of $JE$ and $IE$ at a specific time interval for certain longitudes in three geographic regions (the Eastern and Western Pacific, and the SAA).

All of the values calculated for a specific time within the western longitude sector (Table 3) did not necessarily correspond to a peak value of the local integral flux, which could later be observed. On the basis of the energy flux value we could roughly estimate the energetic electron impact on ionization in TEC units, $Q_{TEC}$, if we used 13.6 eV as the first ionization potential of the oxygen atom and assumed that the decay rate in the topside ionosphere was ~5 x 10$^{-3}$ s$^{-1}$ (see Section 2). By assuming that the conservation of the ionization balance of the $Q_{TEC}$ value was equal to $IE/(13.6 \times 5 \times 10^{-3})$. The calculated $Q_{TEC}$ values for the possible ionization impact were predominately large and ranged from a few TECU to more than one hundred TECU. For comparison, estimated values of TEC enhancements (dVTEC) obtained from the observations are shown within the last column. Only the positive magnitude of the difference between the VTEC values during a storm and during the quiet time were considered here (termed as a positive ionospheric storm). Details for the positive storms of three cases are described in the next section. For most cases, in Tables 2 and 3, estimations of the possible impact of $Q_{TEC}$ exceed the observed residual dVTEC value. As a result, we suggest that extremely intense fluxes of the 10 - 30 keV electrons in the topside low-latitude ionosphere can contribute approximately tens of TECU to localized positive ionospheric storms.

Tables 2 and 3 help indicate the local time dynamics of the electron fluxes. In most cases, electron energy fluxes increased simultaneously (within an hour) for the following two opposite longitudinal sectors: for the Western Pacific (~120°-160°E) during local morning and for the SAA during post-sunset hours, while the fluxes during the daytime (for the Eastern Pacific region) increased one to two hours later. Thus, the enhanced electron energy flux first appeared both within the morning side of the forbidden zone and within the night-time SAA (i.e. the IRB, according to Figure 1 for L-shells of approximately 1.05 and 1.15, respectively). The enhancement then drifted eastward and appeared within the forbidden zone over the Eastern Pacific sector (for simplicity the later times are not shown in Table 3).

## 5. Case events for great electron enhancements

In this section we present a comparative analysis of energetic electron fluxes and ionospheric storm positive phases at low latitudes during major geomagnetic storms on 26 - 27 July 2004, 9 - 10 November 2004, and 14 - 15 December 2006.

### 5.1. Case 1: 26- 27 July 2004

A CME-driven storm from 26 to 27 July 2004 began at ~23 UT on 26 July (see Figure 3). A local peak in storm activity (SYM-H index) was detected at ~02 UT and the main peak was determined at ~12 UT on 27 July [*Zhang et al.*, 2007]. Auroral activity was intense (AE ~ 1500 nT) within the first three hours and moderate (AE ~ 500 nT) during the storm partial recovery phase as a result of the northward turning IMF *Bz* from 02 to 05 UT (labeled 26 to 29 UT in Figure 3a). A strong increase of solar wind dynamic pressures to approximately 30 to 40 nPa between 3 and 6 UT caused compression of the magnetosphere with a decrease of the nose magnetopause distance to ~7 Re as estimated using a model [*Kuznetsov et al.* 1998; *Suvorova et al.*, 1999]. Maximal storm and auroral activity, as seen in the SYM-H and AE indices, was observed several hours later between 08 and 14 UT on 27 July.

*Electron fluxes.* Two hours following storm onset at ~23 UT, during the main and partial recovery phase, three NOAA/POES satellites (P5, P6, and P7) detected great enhancements of quasi-trapped electrons over two energy ranges >30 keV and >100 keV near equatorial latitudes, ±30° (Figure 3 a and b). The great enhancement of >30 keV electrons occurred over a very wide longitudinal range, extending from approximately 30°E through the Pacific to 30°W within the SAA, as clearly seen in Figure 3b against the low background intensity observed during other times over a 36-hour period.

Figure 3a shows time variations for the integral fluxes $F(>E)$ of the >30 keV and >100 keV electrons, as well as for the geomagnetic activity indices SYM-H and AE between 22 UT and 06 UT (labeled 22 - 30 UT). The interplanetary electric field Y component $Ey = Vx·Bz$ (where $Vx$ and $Bz$ are, respectively, the solar wind velocity X-component and the interplanetary magnetic field Z-component) is shown in the panel with the AE index. The >30 keV electron fluxes peaked at approximately $10^7$ units in both hemispheres and maximal fluxes were detected over the Pacific and SAA regions during approximately 5 hours for all local time sectors, while >100 keV electrons had a maximal intensity of ~$10^7$ units only over the area of the night-time SAA. The enhancement of >30 keV electrons within the Eastern Hemisphere (encircled and marked by arrows in Figure 3) were observed by P5, P6, and P7 during ~15 min passes for the low-latitude region at ~0040 UT (90°E; 07LT), ~0220 UT (60°E; 07LT), ~0250 UT (120°E; 10LT), ~0345 UT (160°E; 14LT), and ~0530 UT (130°E; 14LT).

Enhancements are characterized by smooth profiles that indicate non-sporadic electron penetration to the topside ionosphere. The smooth shape and a limited lifetime indicate the gradual and relatively fast transport of electrons within the magnetosphere. The intense fluxes of >30 keV electrons within the forbidden zone (0° - 270° longitudes) persist for a few hours; between 03 UT and 06 UT they peak up to $10^7$ units both on the day and night sides. During this time, geomagnetic activity significantly weakens due to the abrupt northward turning IMF (in Figure 3 the interplanetary electric field, $Ey$, changed to a positive value, respectively) and is followed by a compression of the magnetosphere due to strong increase of the solar wind dynamic pressure.

Interestingly, the >100 keV electron flux was much weaker than that of the >30 keV electrons within the forbidden zone, but the >30 and >100 keV electrons had the same intensities inside of the SAA region (Figure 3a). The >300 keV electrons (not shown) were only enhanced inside of the SAA. Hence, in the energy range from 30 to 300 keV, the spectrum of energetic electrons near the edge of the IRB was descending.

Figure 4 presents the integral energy spectra of electrons at 3 - 4 UT for three different longitudes. Approximations of the spectra by expression (1) are shown by dashed curves. From the approximations, we estimated energy fluxes over the Pacific and SAA regions during the time interval from 03 to 04 UT (Tables 2 and 3). Quasi-trapped electrons within the energy range from 30 to 100 keV produced energy fluxes, $IE$, of ~3.1, 0.2, and 7.1 mW/m$^2$, respectively, at longitudes of 156°E, 132°W, and 78°W. The spectra significantly changed over time in a given longitudinal interval, so, naturally, maximal energy fluxes at different longitudes could occur at different time intervals. For example, we compared fluxes for the Western Hemisphere, at ~0315 UT (132°W) and at ~0500 UT (160°W), see Figure 3a, and found that the energy flux increased from 0.2 mW/m$^2$ to 3.1 mW/m$^2$ within two hours.

A spectrum of low-energy electrons (30 eV to 30 keV), as measured by the DMSP F14 satellite at 0256 UT (~67°E), is shown in Figure 5. The spectrum ascended at energies from 1 to 30 keV. The most intense flux of 3 x $10^7$ units, observed for an energy of 30 keV, and the relevant energy flux density of $JE(<30$ keV$) \sim 7.1·10^{11}$ eV/(cm$^2$ s sr) was equivalent to an $IE$ of ~7.1 mW/m$^2$ (Table 2). Sometimes, a low-energy electron enhancement was accompanied by the enhancement of suprathermal electrons (10-100 eV) of the same order of magnitude (not shown). Although, in this case, POES and DMSP measurements did not match longitudinally; nevertheless, by summarizing two values for low and medium energies, we can approximately estimate the total energy flux within the energy range from 1 - 100 keV in the Eastern Hemisphere as 10 mW/m$^2$. Therefore, the energetic electron impact on the ionosphere-thermosphere system at low latitudes could be very significant.

*Positive ionospheric storm.* Figure 6 shows two-hour GIMs of the residual VTEC (dVTEC) from 20 UT to 06 UT. The residual dVTEC was calculated as the difference between storm and quiet days from 26 to 27 July and 4 August, respectively. A pattern just before the storm onset was mapped to 20 - 22 UT. A small excess ionization of ~10 TECU occurred at low latitudes in the afternoon sector. The positive phase was then gradually amplified during the storm main phase from 23 UT to 02 UT due to the southward turning IMF $Bz$ and the induced dawn-dusk interplanetary electric field (IEF). During the initial phase and the first intensification of the main phase (see maps 22-00 and 0-2 UT), ionization in the crest of the equatorial ionization anomaly (EIA) increased in the afternoon sector over the East Pacific region due to the prompt penetrating electric field (PPEF), as expected from the full electrodynamic scenario [*Fejer*, 2002]. During the partial recovery of the SYM-H (see maps 02-04 and 04-06 UT), a positive storm was unexpectedly maximized particularly from noon to dusk over the Pacific and SAA regions (from ~100°E to 30°W). Also, the VTEC enhancement significantly extended to the mid-latitudes over a wide local time range (from morning to midnight). Therefore, the positive storm at low-to-mid latitudes was intensified in the dayside-dusk-midnight sector under the northward IMF Bz condition, when the mechanism of the PPEF could not operate. Furthermore, a mechanism for the Disturbance Dynamo Electric Field (DDEF) during the recovery stage should result in a depression of the fountain effect in the daytime and the post sunset sectors [e.g., *Fejer*, 2002; *Huang et al.*, 2005; 2010]. Obviously, during nighttime, the DDEF mechanism without an additional ionization source could not produce a significant TEC increase at low-to-mid latitudes.

In this regard, within the same regions and times, enhanced energetic electron fluxes were observed (Figures 3 and 6). The spatial distribution of VTEC enhancements was not uniform and consisted of a number of spots of enhanced ionization. A separate spot at approximately noon in the Southern Hemisphere with a maximum dVTEC of ~35 TECU and a maximum for the >30 keV electron flux of $10^7$ units were observed for the same longitudinal intervals (150° to 160°E) and during the same time (2 to 4 UT). Narrow strips of small increases (<10 TECU) were seen in northern/southern mid-latitudes (~30°) during the morning hours (the encircled regions at 2-4 UT and 4-6 UT in Figure 6), where enhanced particle fluxes of $10^5$ to $10^6$ units were also detected. During the interval from 2 to 6 UT, spots of ionization and concomitant intense fluxes were observed in the East Pacific sector and near the SAA in the afternoon and post-sunset. A sufficiently large dVTEC value of ~10 to 20 TECU and a peak in the electron flux of ~$10^7$ units were observed over the night SAA (the encircled regions in Figure 6). These features cannot be predicted using the full electrodynamic scenario. The EIA crest during local noon was also restricted to within ±15° in geomagnetic latitude during both the main and the partial recovery phases [see *Ngwira et al.*, 2012]. As understood [e.g., *Balan et al.*, 2009], the daytime PPEF together with the equatorward neutral wind should result in an opposite effect, shifting the EIA crest to higher latitudes. After 6 UT the geomagnetic activity again resumed. However, the positive storm weakened rapidly on the dayside, although it persisted after sunset until midnight.

Therefore, we could establish two observable facts that clearly contradicted the electrodynamic scenario - a prenoon-noon and dusk-premidnight positive storm, when the operation of general drivers was tenuous/absent or opposite. As a result, it is quite reasonable to assume that the effect of the neutral wind circulation change due to Joule heating could not play an important role in low-latitude positive storms, particularly at night in the absence of an ionizing source. Also, the effect of equatorward neutral winds should be ruled out during the day by taking into account the sharp change of the solar wind condition at the beginning of the storm. DDEF as main driver during recovery can produce a prominent TEC increase only locally, at the magnetic equator close to sunset; however, this is a seasonal phenomenon that occurs close to an equinox [*Huang et al.*, 2010]. Additionally, the DDEF can strengthen the fountain effect mainly post-midnight. Nevertheless, a positive ionospheric storm lasting for four hours occurred after the first magnetic storm intensification during the pre-midnight period. The combined effect of DDEF/wind and particle ionization could likely result in the wide longitudinal extent of TEC enhancements at low/mid-latitudes. In regards to this ionospheric storm, we note that *Li et al.* [2010] presented interesting observations of F3 layers over Australia and low-latitude ionospheric irregularities over the SAA during the considered time interval from 2 to 4 UT. Such observational features can be also related to particle impacts.

***Electron impact.*** We considered the large enhancement of low- and medium-energy electron fluxes (1-100 keV) as the most likely particle source for abundant ionization. We found that the patterns of the geographic distribution of >30 keV electron enhancements and positive ionospheric storms were very similar (see Figures 3b and 6). The time interval for electron enhancements overlapped with increases in VTEC. If we used a decay rate of the F region density of ~5 x $10^{-3}$ s$^{-1}$, we could roughly estimate an electron impact to ionization, $Q_{TEC}$, of ~29 TECU for 30 to 100 keV electrons and ~65 TECU for 10 to 30 keV electrons, totalling ~94 TECU; a value large enough to supply and maintain the observed ionization increase by ~20 to 35 TECU. As a result, we suggest that energetic electrons produce a significant contribution to the redundant ionization of the ionosphere and, hence, can be considered as an important supplement to other general drivers of the storm-time ionosphere, especially in the morning and night sectors.

*5.2 Case 2: 9-10 November 2004*

The onset of the CME-driven storm on 9 - 10 November 2004 was observed at ~19 UT on 9 November. Storm activity peaked at ~21 UT on 9 November and at ~09 UT on 10 November. The auroral activity was intense within the first two hours (AE < 1500 nT) until 21 UT and weak (AE < 200 nT) during the next six hours after storm partial recovery, caused by the sharp northward turning of the IMF *Bz*. During the partial recovery period, the solar wind dynamic pressure varied strongly [*Balan et al.*, 2008].

***Electron fluxes.*** Figure 7 shows the fluxes of energetic electrons (>30, >100 keV), observed by three NOAA/POES satellites and the geomagnetic activity during the storm. Great enhancements in energetic electrons of ~$10^6$ to $10^7$ units over the Pacific region during the morning (7-10 LT) were observed by P5 and P7 at ~2115 and 2330 UT during periods of weakened auroral activity and storm partial recovery. Intense fluxes of >30 keV and >100 keV electrons within the eastern sector persisted only during two hours from 21 to 23 UT, in the western sector and over the SAA during longer times of 5 to 6 hours.

The characteristics of energetic electrons were very similar to those in Case 1 - short-lived electron enhancements with smooth profiles and a descending spectrum. Figure 8 provides the integral spectra of the electrons at 21 - 22 UT over the Western, Eastern Pacific, and SAA regions. During this time interval, a maximum flux of >30 keV was detected within the Eastern sector at 138°E (Figure 7). We estimated the energy fluxes, *IE*, for 30 to 100 keV as ~3.5, 2.1, and 8 mW/m$^2$ at longitudes of 138°E, 165°W, and 51°W, respectively (Table 2,3).

Examples of the spectra of low-energy electrons measured by the DMSP F14 satellite at approximately 2140 UT and at a longitude of 143°E are presented in Figure 9. The spectrum ascends at energies from 1 to 30 keV. The maximal differential energy flux for low-energy electrons and suprathermal electrons was approximately $10^7$ units at 2140 UT; and the energy flux *JE*(<30 keV) ~ 2·$10^{11}$ eV/(cm$^2$s sr) that provides an *IE* ~2 mW/m$^2$ was integrated over the angle 2π. At 2315 UT and at a longitude of ~159°E, we obtained a *JE*(<30 keV) ~ 2.6·$10^{11}$ eV/(cm$^2$s sr) and a *IE* ~ 2.6 mW/m$^2$. Thus, for the Eastern sector, the total energy flux in the range from 1 to 100 keV could be as large as ~6 mW/m$^2$.

***Positive ionospheric storm.*** Figure 10 shows the residual dVTEC derived from 18 UT to 24 UT on 9 November using a quiet day on 5 November. Two maps for 18 to 20 UT and 20 to 22 UT indicated a development in the positive phase during the afternoon. The VTEC enhancement by ~30 TECU likely relates to the action of the PPEF mechanism during the beginning stage of the storm. Then, after 22 UT, during a partial recovery time, the low-latitude positive storm was significantly amplified and expanded over a wide longitudinal sector of ~300° from the East to the West. The positive storm achieved its maximum between 22 and 02

UT and strongly persisted during the next interval. We noted some features at low latitudes such as numerous local spots of ionization, the occupation of a wide local time area from pre-noon to midnight, and an EIA crest predominately located at lower latitudes that did not expand to middle latitudes. A spot of enhanced ionization at the magnetic equator after sunset was recognized, which would be interpreted as a DDEF effect. However, such a phenomenon is seasonal and occurs during periods close to an equinox [*Huang et al.*, 2010] while the November event was not close to an equinox.

Thus, the geographical distribution of the VTEC increase was not uniform and consisted of a number of spots of enhanced ionization with peaks above 50 TECU. Spots with a maximum dVTEC were observed around noon within the Pacific region and at dusk close to and within the area of the SAA. The spots and encircled enhancements in Figure 10 coincide with regions of intense fluxes of 1 to 100 keV electrons (Figures 7 and 9).

*Electron impact.* As discussed above, from 21 to 24 UT on 9 November intense electron fluxes of $\sim 10^6$ to $10^8$ units were observed over a wide longitudinal range from the West Pacific to the SAA regions (Figure 7). In a similar manner to Case 1, the particle and ionization patterns of the geographic distribution and the temporal dynamics well coincided (Figures 7 and 10). Rough estimations for the electron impact, $Q_{TEC}$, over the energy range from 1 to 100 keV were $\sim 51$ TECU against the observed value of $\sim 15$ TECU within the Eastern sector during local morning; and $\sim 20$ and $\sim 74$ TECU within the Western sector against the observed value of $\sim 30$ to 50 TECU at noon and dusk (Table 2, 3). Hence, Case 2 also supported an assumption regarding prominent energetic electrons contribution to the redundant ionization of the ionosphere.

*5.3 Case 3: 14-15 December 2006*

As for the previous two storms, the major geomagnetic storm from 14 to 15 December 2006 was also ascribed to the CME-driven event [*Liu et al.*, 2008; *de Jesus et al.*, 2010]. The storm initial phase began at approximately 14 UT on 14 December. A strong negative IMF $Bz$ with an $\sim$8-hour duration caused a storm onset at $\sim$23 UT with a deep main phase and a prolonged minimum (SYM-H $\leq$ -200 nT) (see Figure 11). The auroral activity was very intense during 24 hours, with peak values of AE > 2000 nT. The magnetosphere suffered the strongest compression resulting from solar wind dynamic pressure during the beginning stage of the storm between 20 and $\sim$24 UT. Earlier we analyzed the simultaneous observations of energetic electrons and a positive storm that was mainly focused in eastern longitudes near noon [*Suvorova et al.*, 2012]. Here we present an extended analysis for different longitudes and local times.

*Electron fluxes and positive storms* Figure 11 (middle panel) displays the pattern for the geographical distribution of >30 keV electron fluxes accumulated for 33 hours during the orbits of five NOAA/POES satellites (P2, P5, P6, P7, and P8). The most intense electron fluxes (>$10^6$ units) were observed between 2 and 6 UT on 15 December. In the bottom panel of Figure 11, a map of the differential dVTEC for the 2 - 4 UT interval is presented as an example of the positive ionospheric storm phase with a maximal value of $\sim$40 TECU. In Tables 2 and 3 one can see that the energy fluxes, *IE*, at different locations and local times were relatively moderate - 1.4 mW/m$^2$ (POES) at noon in the Western Pacific sector; $\sim$0.4 mW/m$^2$ (DMSP) in the morning over Africa; and $\sim$0.6 mW/m$^2$ at dusk in the Eastern Pacific sector and during pre-midnight in the SAA. As a result, the particle impact, $Q_{TEC}$, could be estimated to be approximately 6 and 17 TECU during the night and daytime, respectively.

*COSMIC/FORMOSAT-3 RO-images.* We also analyzed the EC height profiles of the electron concentration using ionospheric tomography produced by COSMIC/FORMOSAT-3. Figure 12 presents pairs of the storm-time and quiet-time meridional cuts for the EC at 2 - 4 and 4 - 6 UT for the following longitudes: 60°E, 120°E, 150°E, 160°W, 105°W, and 75°W that spanned the local time from morning to pre-midnight. Important to point out is that EC enhancements expand significantly to higher altitudes (up to 600 km and above) those seen at all longitudes over the Pacific region and the SAA. A similar pattern was revealed during the entire main phase and during the maximum of the geomagnetic storm from 00 to 06 UT on 15 December. An asymmetry in the height of the F-layer bottom edge was determined. Lowering was determined at the northern latitudes, in contrast to the raising at the southern latitudes (see, for example, the maps in Figure 12a for 120°E, 150°E, and 160°W). Another feature of the storm-time EC was a wide latitudinal extent. We also noted a particularly significant density enhancement at the northern mid-latitudes at 2 - 4 UT. One can clearly see a prominent uplift of the F-region during all local times. In particular, the uplift seemed improbable on the night side (19 - 21 and 21 - 23 LT). Such a nighttime feature was observed close to the SAA (105°W and 75°W). In general, the elevation of the EC to higher altitudes during the daytime equatorial region proves the presence of a strong dawn-dusk electric field (PPEF) operating together with equatorward neutral winds from higher latitudes [e.g., *Balan et al.*, 2009]. However, in the evening-premidnight sector the mechanisms of PPEF, DDEF, and disturbed neutral winds cannot explain long-lasting prominent enhancements over a wide latitude range ± 50° from 200 to 600 km without involving an additional ionization source.

***HORT and LORT images (beacon method).*** Figure 13 displays maps for the vertical TEC for different time intervals on 15 December 2006. The maps resulted from HORT reconstructions. During the storm periods, the EIA was observed as a multi-peak spotty structure with complex dynamics. Typically, spots with a size of a few hundreds of kilometers are not resolved with GIM maps. In the most cases, spots on HORT maps are not artifacts. If the corresponding HO satellites have passed through the region of study at the time of reconstruction, the raw HORT data contain indications confirming that the spots imaged by HORT reflect real anomalies within the electron density distribution.

LORT cross sections (see Figure 14) displayed the marginal portion of the equatorial anomaly (14° - 18°N latitudes), with a core oriented along the magnetic field and wavelike disturbances observed to the north of the EIA crest. The reconstruction also indicated that the main maximum of the EIA crest was elevated to a height of 350 km (south of the image) as compared to 250 km within the northern segment of the reconstruction. Similar features at 120°E were seen in the RO-image from COSMIS/FORMOSAT-3 data (Figure 12a). The core of the anomaly in the LORT image corresponded to the edge of the EIA crest in the HORT image above the Philippines when the FM1 satellite flew over the edge of the anomaly above the Philippines (map of TEC at 0400UT in Figure 13). Here, it should be noted that on 15 December 2006, the northern crest of the EIA persisted a long time from 02 to 17 UT, and individual fragments of the crest existed up to 21 UT.

The examples provided in Figures 14b and c present the LORT images, respectively, for 0645 UT and 0900 UT which display the crest of the EIA with characteristic structural features - the well-matured core of the anomaly was oriented in the direction of the geomagnetic field; the equatorial anomaly and variations in the thickness of the ionospheric layer were distinctly asymmetric. The map for the TEC at ~07 UT (Figure 13) displays the EIA over the region of Taiwan. The map of the TEC at ~09 UT (Figure 13) also clearly indicates the EIA crest south of Taiwan (in the interval from 18° - 20°N ).

Therefore, results of the ionospheric density reconstruction using the various methods (GIM, RO, HORT, and LORT) were all in good agreement. RT-imaging revealed a complex structure of the positive storm at the low/mid-latitudes during a storm maximum between 2 and 6 UT. We believe that the observed specific features, the spatial inhomogeneity, small-scale spots, wavelike structures, nighttime F-region uplifting, etc. are evidence of an energetic electron impact in the topside ionosphere.

## 6. Discussion

We studied the relationship between energetic quasi-trapped electrons and ionospheric ionization for major geomagnetic storms during the period from 1999 to 2006. We analyzed storm-time events for times when the inner radiation belt approached heights of the topside low-latitude ionosphere. During such events, a dramatic increase in particle flux of a few orders of magnitude relative to the pre-storm level was observed. In this work we focused on the West Pacific region where salient events rarely occur.

A comparison of the patterns between the ionospheric storm positive phase and quasi-trapped >30 keV electron fluxes revealed good spatial and temporal coincidences. Low/mid-latitude positive ionospheric storms, concomitant with intense equatorial electron fluxes, achieved large magnitudes of 10-30 TECU in the morning and 35-50 TECU in the post-sunset sectors. As examples, we considered geomagnetic storms on 26-27 July, 9-10 November 2004, and 14-15 December 2006. Due to the sequence of coronal mass ejections, a similarity between the two storms during 2004 indicated that both were of the super-storm series. Electron fluxes during other storm activations for each super-storm series were moderate or weak. The storm during December 2006 was isolated. An analysis of this storm was performed using various ionospheric measurement techniques.

The following major storms have attracted attention to date because they resulted in a specific thermosphere-ionosphere-magnetosphere system response: the July event [*Burke et al.*, 2007; *Kunitsyn et al.*, 2008a; *Lazutin*, 2012; *Li et al.*, 2010; *Ngwira et al.*, 2012; *Pedatella et al.*, 2008; *Stephan et al.*, 2008]; the November event [*Fejer et al.*, 2007; *Paznukhov et al.*, 2007; *Balan et al.*, 2008; 2009; *Huang*, 2008; *Mannucci et al.*, 2009; *Deng et al.*, 2009; *Sahai et al.*, 2009; *Kelley et al.*, 2010; special issue of JASTP, 2010; *Solovyev et al.*, 2011; *Abdu*, 2012]; and the December event [*Lei et al.*, 2008a; 2008b; *Dmitriev et al.*, 2008; 2010; *Pedatella et al.*, 2009; *de Jesus et al.*, 2010; *Klimenko and Klimenko*, 2012; *Wei et al.*, 2011; *Suvorova et al.*, 2012].

From the studies indicated above, we summarize the following interesting features observed during the same spatial and temporal intervals as for those analyzed in our paper: unusual uplifting of the F region [*de Jesus et al.*, 2010; *Ngwira et al.*, 2012; *Abdu*, 2012]; the development of a strong F3 at the topside ionosphere for different local times during the low-occurrence season [*Paznukhov et al.*, 2007; *Balan et al.*, 2008; *Li et al.*, 2010; *Kelley et al.*, 2010]; enhanced ionospheric scintillation activity and a large longitude extent for ionospheric irregularities during the low-occurrence season [*Sahai et al.*, 2009; *de Jesus et al.*, 2010; *Li et al.*, 2010; *Kelley et al.*, 2010; *Abdu*, 2012]; the suprathermal particle effect at mid-latitudes [*Pedatella et al.*, 2009; *Ngwira et al.*, 2012]; and the fast rebuilding of the radiation belt [*Lazutin*, 2012]. Some studies have indicated specific features for storms. For example, for the 9 November storm, *Paznukhov et al.*, [2007] emphasized that the positive storm reached a maximum during the recovery phase when the daytime equatorial electric field originated from the west; *Mannucci et al.* [2009] found unusual early morning enhancements in the F layer during the recovery phase (as for the October 2003 storm, see *Batista et al.* [2006]); and for the 27 July storm *Pedatella et al.* [2008] found large changes in the longitudinal structure for the low-latitude ionosphere. The researchers applied general mechanisms involving PPEF, DDEF, and equatorward neutral wind effects for the interpretation and modeling of unusually long-lasting positive storms [e.g, *Fejer et al.*, 2007; *Huang et al.*, 2008; *Balan et al.*, 2009; *Mannucci et al.*, 2009; *Kelley et al.*, 2010; *Sahai et al.*, 2009]. The results of the investigations pose challenges to the standard mechanisms identified for positive storms.

On the base of statistics and a case-event analysis, we determined that magnetospheric particle phenomena near the equator can play an important role in long-duration positive ionospheric storms. In this work we have demonstrated that the energy flux of quasi-trapped >30 keV electrons can be large enough to significantly contribute to the ionization of the topside low-latitude ionosphere.

Here, it should be noted that POES electron measurements can be corrupted by proton contamination [*Evans and Greer*, 2004]. Electron detectors are sensitive to >210 keV protons, which are able to pass through the passive shields of detectors. As a result of radiation damage of silicon solid-state detectors in proton telescopes (the so-called aging effect), the correction of proton contamination is difficult. Therefore, the following problems exist: 1) the removal of the contribution of protons from electron counts, and 2) the aging effect of proton telescopes. The first problem can be resolved with a method suggested by *Rodger et al.* [2010] who quantified the level of contamination for "good" electron data (i.e. the counts are most likely to be dominated by electrons). Electron counts should be at least twice as large as the counts from proton detectors. We found that during electron enhancement events, proton fluxes with energies >80 keV and >240 keV at L <1.15 remained smaller by several orders of magnitude for all longitudes including for the SAA region. Therefore, the contribution of energetic protons to electron counts at low latitudes was negligibly small for all of the events considered. The second problem arose because satellites frequently pass over the SAA where the inner energetic proton belt is lower and the proton flux is always large. The proton detector suffers from an aging effect that becomes significant after 2-3 years of operation and that results in errors in proton counts. *Dmitriev et al.* [2010] investigated this effect for NOAA/POES-15, 16, and 17 measurements at the end of 2006 by comparing measurements from new instruments onboard METOP-2 and found that the underestimation of energetic proton fluxes did not exceed 20%. The change was small enough in comparison to the observed electron flux increases. Hence, during the time period before 2006 the effect could be neglected.

Comparing the energy flux of energetic electrons with other sources of ionospheric ionization such as the solar EUV irradiance, solar X-flares, and plasmaspheric plasma fluxes is useful. Under non-flare conditions, the ionospheric response to solar EUV irradiance variability is 7 - 15 TECU per 1 mW/m$^2$ [*Lean et al.*, 2011a; 2011b]. The energy flux of solar irradiance during an X17 class solar flare on 28 October 2003 increased from the pre-flare level of ~4 mW/m$^2$ to ~13 mW/m$^2$ [*Strickland et al.*, 2007] and produced an ionospheric enhancement of ~25 TECU at the subsolar point that was 30% above background [*Tsurutani et al.*, 2005].

The variability in the TEC measured by ground-based GPS stations is also related to plasmasphere-ionosphere coupling. The plasmasphere is essentially an extension of the ionosphere and begins at an altitude of ~1000 km where hydrogen ions replace oxygen ions and become the main plasma component. That the plasmasphere plays a significant role in the maintenance of the nighttime ionosphere due to the downward plasma flux is well-known, while on the dayside the plasmasphere is filled by upward plasma fluxes from the ionosphere. Hence, the plasmasphere certainly contributes to ground-based GPS TEC measurements (see *Pierrard et al.* [2009] for review). Although the plasma flux strongly depends on solar activity, season, and geomagnetic activity, its order of magnitude is only $10^8$ cm$^{-2}$s$^{-1}$. For example, under quiet conditions the downward plasmaspheric flux into the nighttime ionosphere, on average, is ~2 x $10^8$ cm$^{-2}$s$^{-1}$. Hence, the plasmaspheric contribution to the total electron content is typically a few TECU (2–4 x $10^{12}$ el/cm$^2$). A maximum effect of approximately 5-6 TECU was observed at 2 - 4 LT and 14 - 16 LT in the low-latitudinal region, while at other local times and higher latitudes the value did not exceed 2 TECU [*Yizengaw et al.*, 2008]. Since the plasmasphere is depleted during a storm, its contribution also decreases. Therefore, if expected particle-induced TEC enhancements at a given local time during major storms exceed the plasmaspheric TEC value by at least two-three times, the contribution of particle ionization is meaningful.

Another important issue is the conditions required for the downward transport of electrons from the IRB to heights below 1,000 km. A well-known mechanism is radial diffusion across drift shells. However, for strong magnetic field at low altitudes, diffusion is very slow and results in very weak fluxes of energetic electrons for forbidden drift shells. In contrast, geomagnetic storms, large substorms, and sudden commencements or magnetospheric compressions are accompanied by a very strong penetrating electric field (EF) in the dawn-dusk direction of approximately a few mV/m [e.g., *Nishimura et al.*, 2006; 2009; *Shinbori et al.*, 2006; *Fejer et al.*, 2007; *Lazutin and Kuznetsov*, 2008]. On the nightside, this EF is pointed westward and results in the fast (a few hours) *ExB* drift of particles across magnetic field lines toward the Earth. Unfortunately, the storm-time generation of EFs within the inner magnetosphere is poorly studied.

*ExB* drift must produce the effective transport of high-energy radiation belt electrons as well [e.g., *Lazutin and Kozelova*, 2012]. Therefore, the 0.3 to 1.5, and 1.7 to 3.4 MeV electron fluxes measured onboard the SERVIS-1 satellite were inspected. Figure 15 shows SERVIS-1 data on electron fluxes at low *L*-shells for the magnetic storm on 9 November 2004. During the storm interval we observed an increase in high-energy electron fluxes that occurred during the same time as the 30 keV electron event described above. The same situation was found for the 27 July 2004 magnetic storm. Hence, electrons over a wide energy range from 30 keV to >1.7 MeV appeared at low *L*-shells so simultaneously that our idea regarding earthward electron transport due to *ExB* drift, which is energy independent, was supported.

Very strong impulses of the induced EF of ~20 mV/m were revealed in the mid-latitude magnetosphere [e.g., *Kozelova et al.*, 2000]. Likely, similar strong impulses of EF can penetrate to low latitudes at heights of lower edge of the IRB. If so, electrons of the RB will suffer *ExB* drift toward the Earth. During quiet-times, the energy spectrum of the trapped electron population at $L \sim 1.23–1.34$ has a local peak at 30 keV [e.g., *Kudela et al.*, 1992]. During magnetic storms, these electrons suffer *ExB* drift toward Earth that results in the largest flux enhancements. Then electrons drift azimuthally toward the east along drift shells whose altitude above the Pacific region decreases with an increasing longitude. In the morning sector, energetic electrons can reach and ionize the topside ionosphere. The ionization effect for the intense flux of energetic electrons at heights of ~400 to 800 km has been estimated to be ~20 TECU that is equivalent to the formation of the storm-time ionospheric F3 layer. As mentioned above, the appearance of the F3 layer during July and November 2004 events have been reported by several authors. Therefore, the direct ionization of the topside ionosphere by quasi-trapped energetic electrons can be considered as an important contribution to low-latitude positive ionospheric storms, particularly during morning and night hours.

## 7. Conclusions

From a statistical and case-event analysis of major geomagnetic storms accompanied by magnetospheric particle phenomenon, we found that energetic electron enhancements are an important source of ionization in the topside ionosphere. Although the particle phenomenon has a certain relationship to the geomagnetically disturbed state of the magnetosphere, we did not determine a dependence on the geomagnetic storm magnitude. For nine salient cases we demonstrated that the positive phase of ionospheric storms observed over the Pacific in the morning and pre-midnight hours, particularly during storm recovery, can be explained by the direct ionization produced by the intense flux of quasi-trapped energetic electrons in the topside ionosphere rather than by PPEF, DDEF, and/or equatorward neutral winds only. Therefore, we suggest that the "auroral"-level intensity of quasi-trapped electrons with energy of 10 to 30 keV can be considered as an important supplement to general ionospheric drivers.

In summary, during magnetic storms energetic (~30 keV) electrons drift fast radially from the IRB to ionospheric altitudes located within the nightside sector. Then, quasi-trapped electrons, drifting azimuthally eastward, exhaust their energy during the ionization of atmospheric gases and produce abundant ionization within the low- and mid-latitude ionosphere.

The phenomena of enhanced quasi-trapped energetic electron fluxes are of great interest to researchers of ionospheric-magnetospheric coupling at low and mid-latitudes. Therefore, new results from studies of enhanced, quasi-trapped energetic electron fluxes and the coupling of particle phenomena to TEC increases and uplifts of the F region allow a view of the problem of long-duration positive ionospheric storms from a new perspective.

**Acknowledgements**
The authors thank a team of NOAA's Polar Orbiting Environmental Satellites for providing experimental data about energetic particles and Kyoto World Data Center for Geomagnetism (http://wdc.kugi.kyoto-u.ac.jp/index.html) for providing the geomagnetic indices. Authors are grateful to Professor N. Hasebe from Waseda University, Japan, for the SERVIS-1 data. The GIM data were obtained through ftp://ftp.unibe.ch/aiub/CODE/. The ACE solar wind data were provided by N. Ness and D.J. McComas through the CDAWeb website. The DMSP particle detectors were designed by Dave Hardy of AFRL. JHU/APL provided the DMSP particle spectrogram, which is made possible by P. T. Newell and T. Sotirelis. The authors thank C.-M. Huang and M. Schulz for helpful discussions. This work was supported by grants NSC 100-2811-M-008-062 and NSC 100-2119-M-008 -019.

**Table 1.** List of the Storms with Great Electron Enhancements.

| Storm day, main phase time | Quiet day beginning | Minimum of $Dst$ (nT) | Storm phase* |
|---|---|---|---|
| 15 July 2000 19-22 UT | 2 July, 0 UT | -340 | main |
| 29-30 October 2003 6-9; 18-02 UT | 11 October, 06 UT | -400 | ini, main, part.recovery |
| 30 October 2003 20-23 UT | 11 October, 06 UT | -400 | ini, main, recovery |
| 26-27 July 2004 23-02; 5-12 UT | 3 August, 12 UT | -200 | ini, main, part.recovery |
| 9-10 November 2004 19-21; 4-10 UT | 5 November, 6 UT | -250 | part.recovery |
| 21 January 2005 19-21 UT | 5 February, 0 UT | -100 | ini, main |
| 15 May 2005 6-8 UT | 26 May, 0 UT | -300 | recovery |
| 12-13 June 2005 17-00 UT | 21 June, 16 UT | -120 | main |
| 14-15 December 2006 23-01 UT | 4 December, 10 UT | -200 | recovery |

*Storm phase associated with enhancements of the electron fluxes in the Eastern hemisphere.

**Table 2.** Estimate of Energy Deposition of Quasi-Trapped Electrons (30-100 keV) and Observed dVTEC in the Eastern hemisphere.[a]

| Date & time | Ns[b] | Lon_E/LT [c] °/hours | JE[d] ×10$^{12}$ | IE, mW/m$^2$ | Q$_{TEC}$, TECU | dVTEC, TECU |
|---|---|---|---|---|---|---|
| 15 July 2000 21-23 UT | 1 4 | 153°/7 132°/6 | 0.053 0.08 0.13 | ~0.5 0.8 1.3 | ~5 7.4 ~12.2 | 10-20 5-15 |
| 29 October 2003 21-22 UT | 3 4 | 144°/7 150°/7 | 0.5 0.15 0.65 | 5 1.5 6.5 | 46 13.8 ~60 | 5-12 5-12 |
| 30 October 2003 20-21 UT | 3 4 | 150°/6 151°/6 | 1.13 0.51 1.64 | 11.3 5.1 16.4 | 104 47 148 | 5-30 5-30 |
| 27 July 2004 3-4 UT | 3 4 | 156°/14 68°/8 | 0.31 0.71 1.02 | 3.1 7.1 10.2 | 28.5 65.3 ~94 | 20-35 5-12 |
| 9 November 2004 21-22 UT | 3 4 | 138°/6 143°/6 | 0.35 0.2 0.55 | 3.5 2 5.5 | ~32 18.4 ~51 | 10-15 10-15 |
| 21 January 2005 20-21 UT | 3 4 | 144°/5 176°/8 | 0.049 0.45 0.5 | ~0.5 4.5 5 | 4.6 ~41.6 ~46 | 3-7 10-20 |
| 15 May 2005 9-11 UT | 3 4 | 165°/20 130°/18 | 0.5 0.04 0.54 | 5 0.4 5.4 | 46 ~3.7 ~50 | 15-30 30-45 |
| 12 June 2005 19-21 UT | 4 4 | 159°/6 147°/5 | 0.054 0.12 0.17 | 0.5 1.2 1.7 | ~5 11 16 | 4-7 3-5 |
| 15 December 2006 2-4 UT | 5 4 | 166°/13 54°/6 | 0.143 0.04 0.18 | 1.4 0.4 1.8 | ~13 3.7 ~17 | 20-40 3-5 |

[a] Two lines in columns #2-6 represent the POES and DMSP data, a third line is a total value.

[b] Number of POES and DMSP satellites

[c] Eastern longitude and local time

[d] Value of JE is in eV/(cm$^2$ s sr).

**Table 3.** Estimate of Energy Deposition of Quasi-Trapped Electrons (30-100 keV) and Observed dVTEC in the Western hemisphere.[a]

| Date & storm time | Lon_W/LT[b], °/hours | $JE^c \times 10^{12}$ | IE, mW/m² | $Q_{TEC}$, TECU | dVTEC, TECU |
|---|---|---|---|---|---|
| 15 July 2000 21-23 UT | - 36°/19 | 0.27 | 2.7 | - ~25 | - 25-50 |
| 29 October 2003 21-22 UT | 159°/10 45°/19 | 0.18 0.25 | 1.8 2.5 | ~17 23 | 20-50 20-50 |
| 30 October 2003 20-22 UT | 153°/10 39°/19 | 0.015 0.1 | 0.15 1 | ~1.4 ~10 | 20-50 20-50 |
| 27 July 2004 3-4 UT | 132°/19 78°/22 | 0.018 0.71 | 0.18 7.1 | ~1.7 ~65 | 15-35 20-35 |
| 9 November 2004 21-22 UT | 165°/10 51°/19 | 0.21 0.8 | 2.1 8 | ~20 ~74 | 30-50 20-35 |
| 21 January 2005 20-21 UT | 150°/10 45°/18 | 0.29 0.96 | 2.9 9.6 | ~27 ~89 | 15-35 15-35 |
| 15 May 2005 9-11 UT | 129°/2 75°/6 | 0.17 0.33 | 1.7 3.3 | ~15 30 | 5-7 <0 |
| 12 June 2005 19-21 UT | 147°/10 75°/15 | 0.045 0.48 | 0.45 4.8 | ~4 44 | 5-15 10-15 |
| 15 December 2006 2-4 UT | 156°/17 69°/22 | 0.064 0.056 | 0.64 0.56 | ~6 ~5 | 10-40 ~7-12 |

[a] Only data of the POES satellites are used. Two lines in columns #2-6 represent data at longitudes in the Pacific and SAA regions.

[b] Western longitude and local time

[c] Value of JE is in eV/(cm² s sr) .

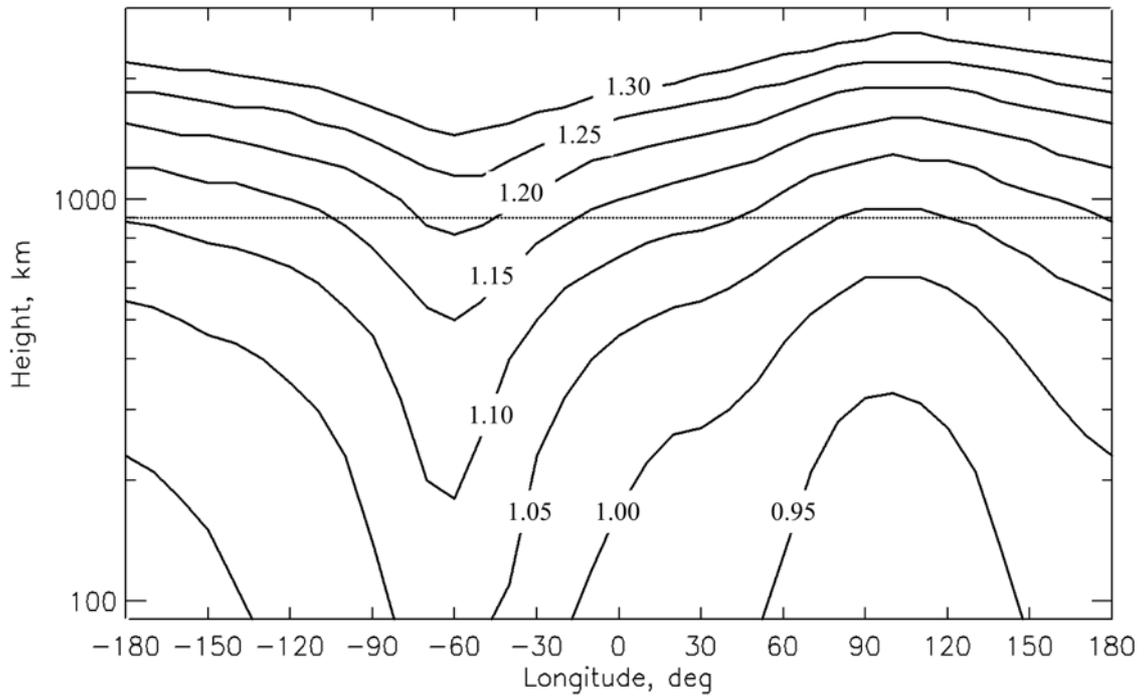

Figure 1: Longitudinal variation of the height of various drift shells at geomagnetic equator calculated from IGRF model for epoch of 2005. Quasi-trapped energetic electrons drift eastward along the drift shells and pass the highest (lowest) heights in the Indochina (SAA) region. Horizontal dashed lines indicate the heights of 300 km and 900 km.

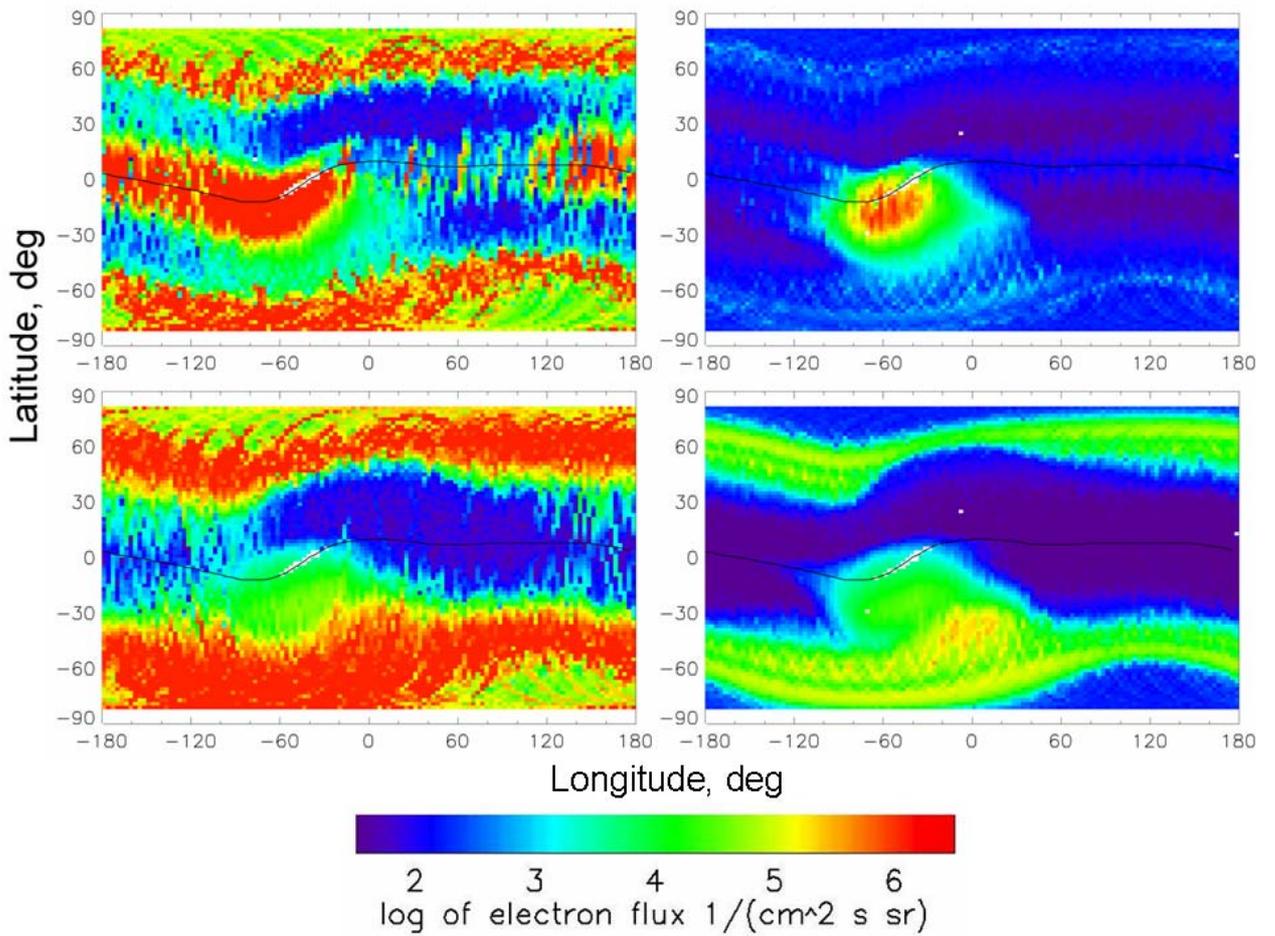

Figure 2. Geographic distributions of the >30 keV electron fluxes from the MEPED instrument: (upper panels) 0º-telescope and (lower panels) 90º-telescope. These maps are composed from data retrievals over multiple orbits of the NOAA/POES satellites at 850 km altitude during (left) storm days and (right) quiet days listed in Table 1. Dip equator is indicated by black curve. Local time of some intense fluxes at low-latitudes is indicated in the Tables 2 and 3 (see column 3).

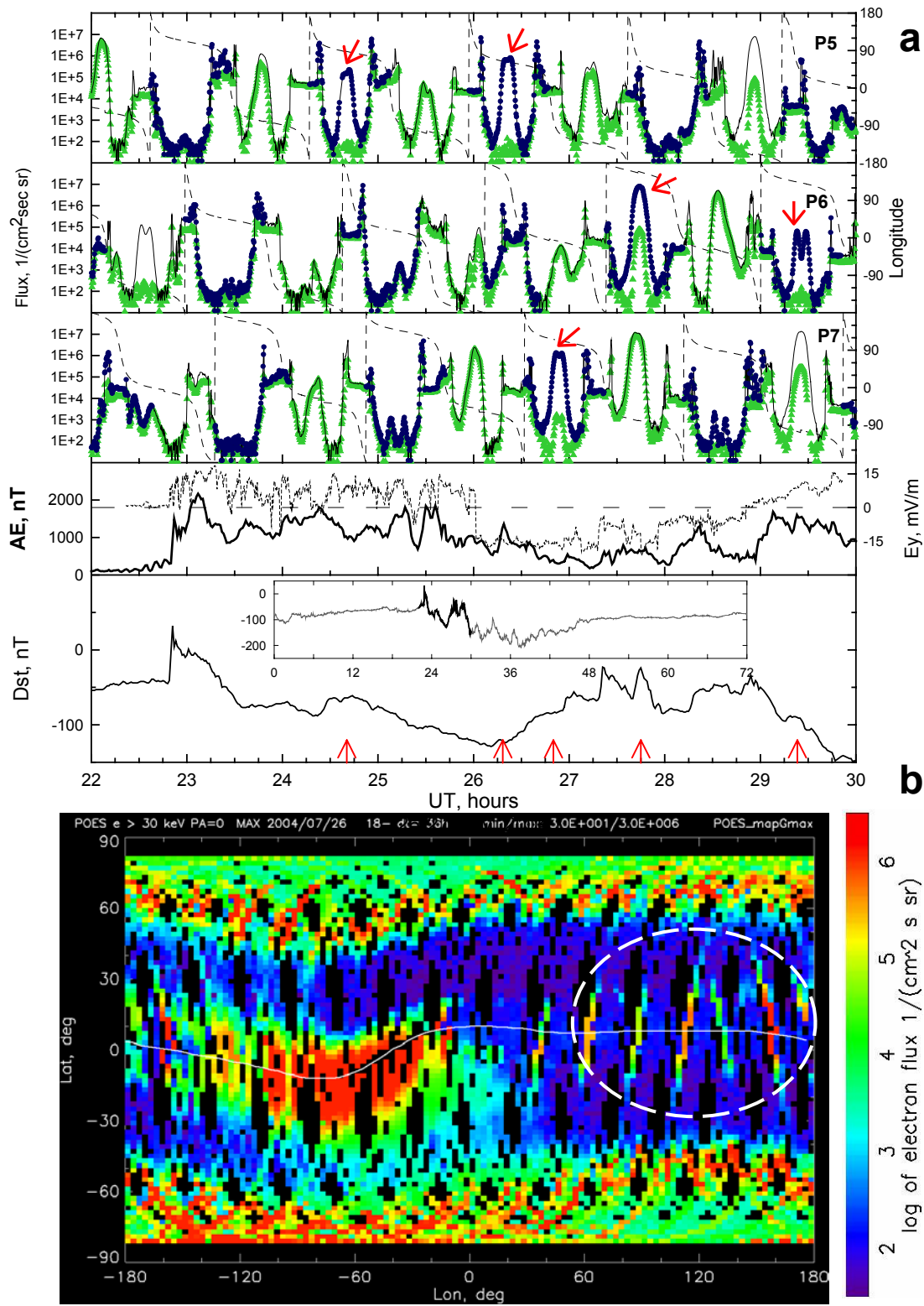

Figure 3. Great electron enhancements during the geomagnetic storm on 26-27 July 2004. (a) From top to bottom: three panels with time profile of the fluxes of the >30 keV (solid curves and dark dotted segments for the Eastern longitudes) and >100 keV electrons (green triangles) observed by NOAA-15 (P5), NOAA-16 (P6), and NOAA-17 (P7) and the geographic longitudes (dashed curves) along the satellite orbital passes. The great electron enhancements at the Eastern longitudes are pointed by red arrows. The AE index and the interplanetary electric field Ey

(dawn-to-dusk) measured by ACE upstream monitor with ~22 min delay. The SYM-H index during 22 UT- 06(30) UT, an inset shows the SYM-H index during three days of 26-28 July, the analyzed time interval is indicated by darker color. (b) Geographic map of storm-time >30 keV quasi-trapped electron fluxes over a 36-hour period beginning from 18 UT, 26 July. Encircled electron enhancements correspond to arrow-marked fluxes in upper panels (a) between 0 UT and 6 UT. The solid white curve indicates the dip equator.

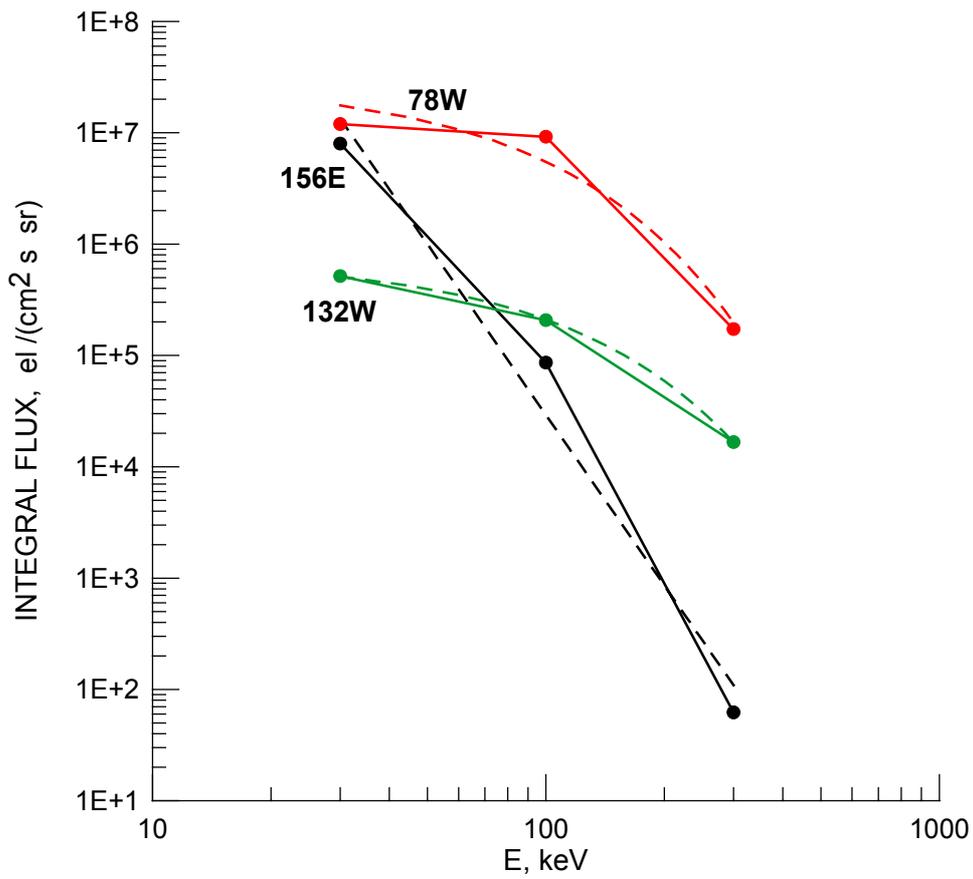

Figure 4. Integral energy flux spectra of the quasi-trapped energetic electrons measured by NOAA/POES satellites on 27 July 2004 at 3-4 UT in three geographic regions: Western Pacific (156°E), Eastern Pacific (132°W) and SAA (78°W). Dashed curves are approximations by power or exponential laws (see expressions (1)).

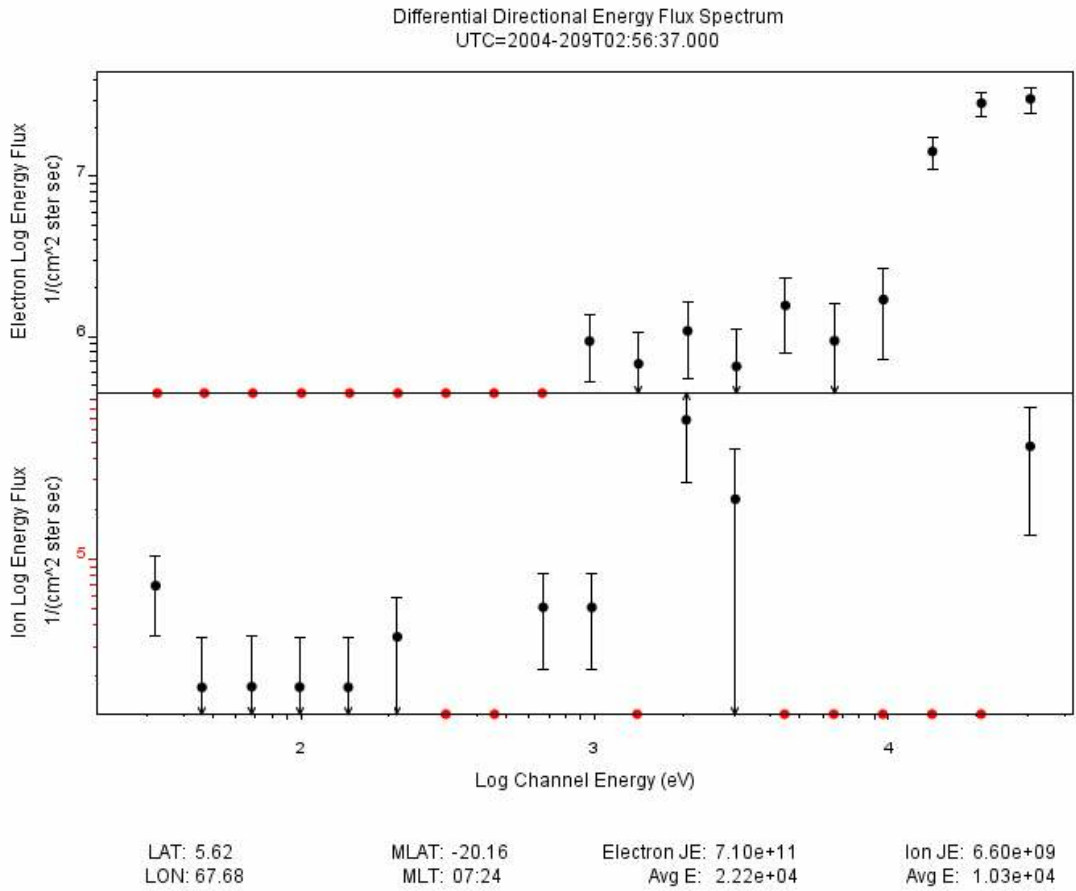

Figure 5. Differential energy flux spectrum measured by DMSP 14 around 3 UT on 27 July 2004 at ~68°E near the equator.

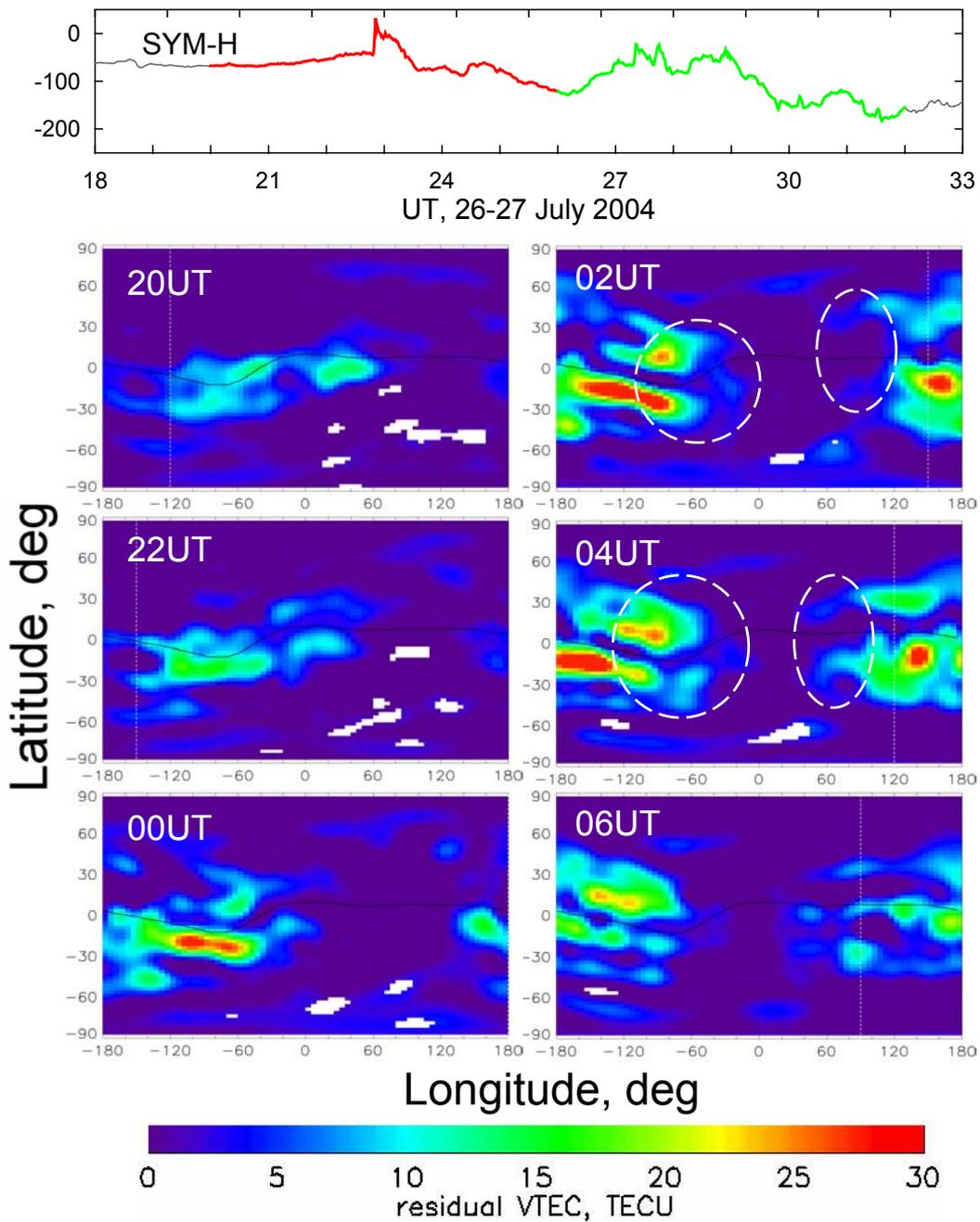

Figure 6. Ionospheric storm positive phase during the geomagnetic storm on 26-27 July 2004. Upper panel: the SYM-H index from 18 UT, 26 July to 9 UT, 27 July. Red and green segments conventionally show the storm initial phase with 1st intensification and the partial recovery with 2nd intensification. Lower panels: 2-hours geographic ionospheric maps of positive phase calculated as VTEC differential between storm-time and quiet-time GIMs, a color pattern corresponds to positive residuals. Dip equator is indicated by black curve. Dashed white line is local noon at the beginning of 2-hour interval.

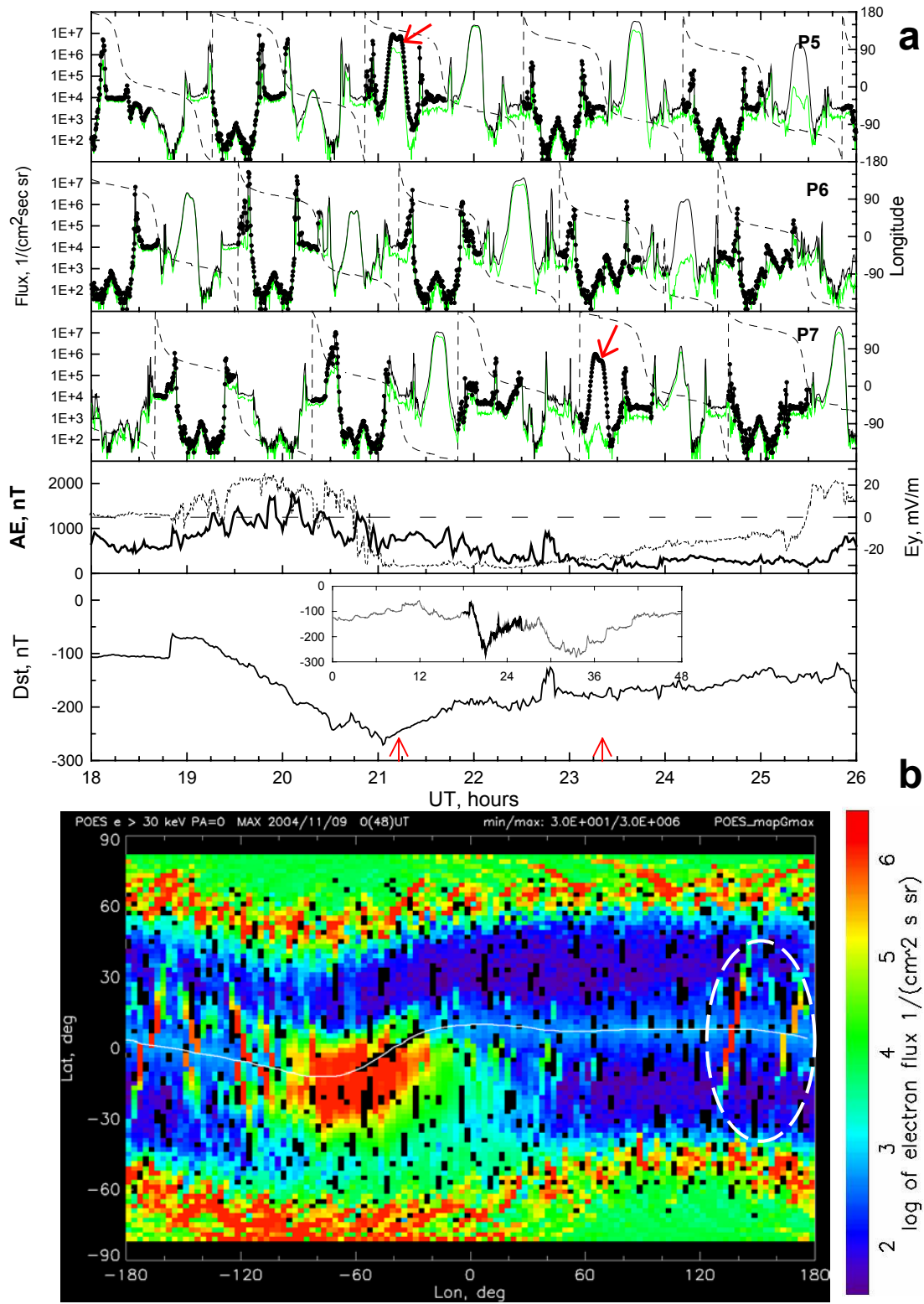

Figure 7. Same as Figure 3 but for 9-10 November 2004, 18 UT-02(26) UT. (a) The IEF Ey from ACE is delayed by ~30 min, an inset in the 'Dst'-panel shows the SYM-H index during period of two days 9-10 November. (b) The map is accumulating data over a 48-hours period beginning from 0 UT, 9 November. Encircled electron enhancements correspond to the arrow-marked fluxes in upper panels (a) between 21 UT and 24 UT.

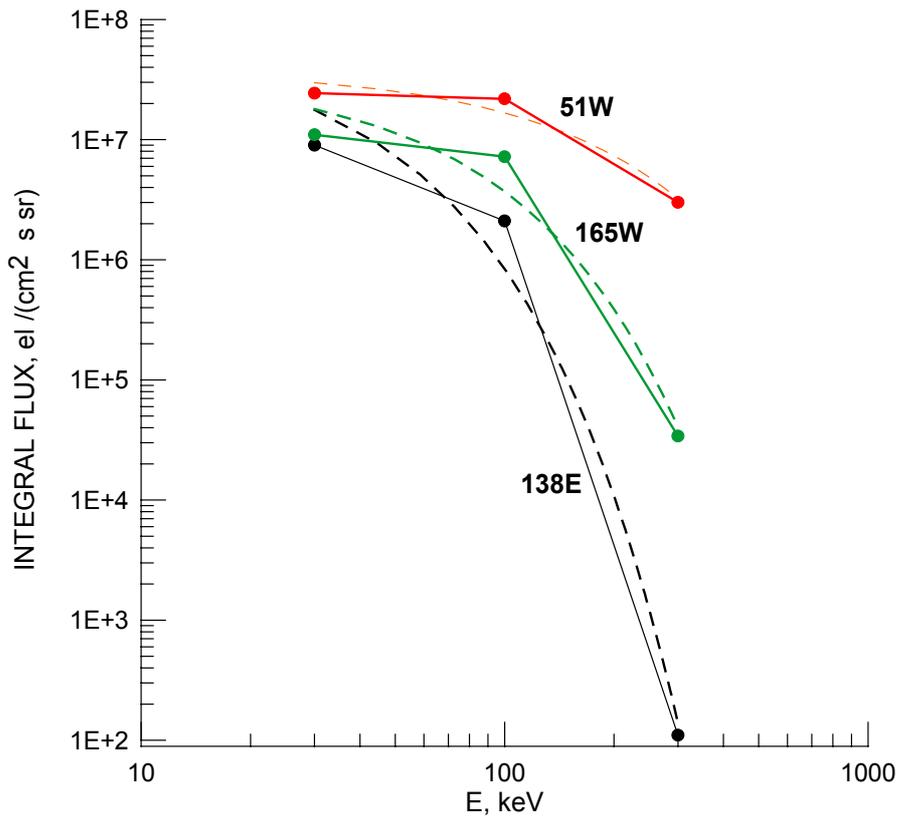

Figure 8. Same as Figure 5 but for 9-10 November 2004 at 21-22 UT.

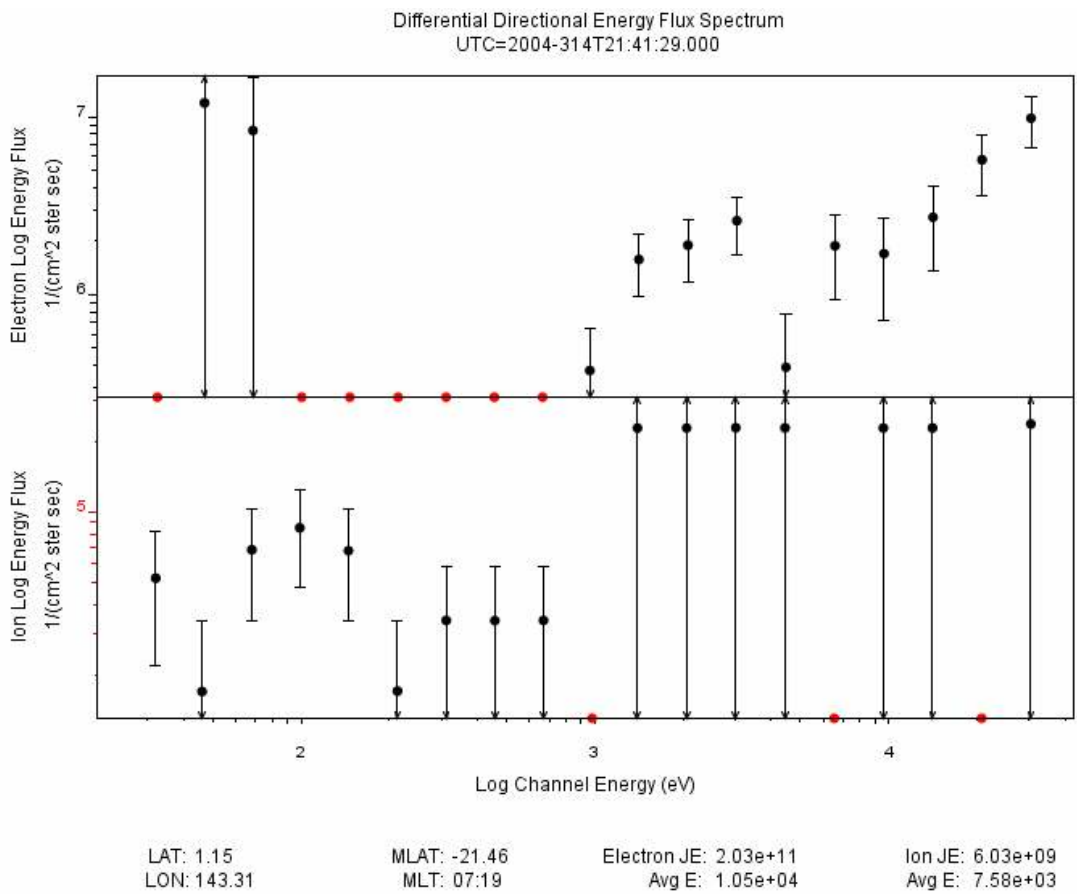

Figure 9. Differential energy flux spectrum measured by DMSP 14 around 21:40 UT on 09 November 2004 at 143°E near the equator.

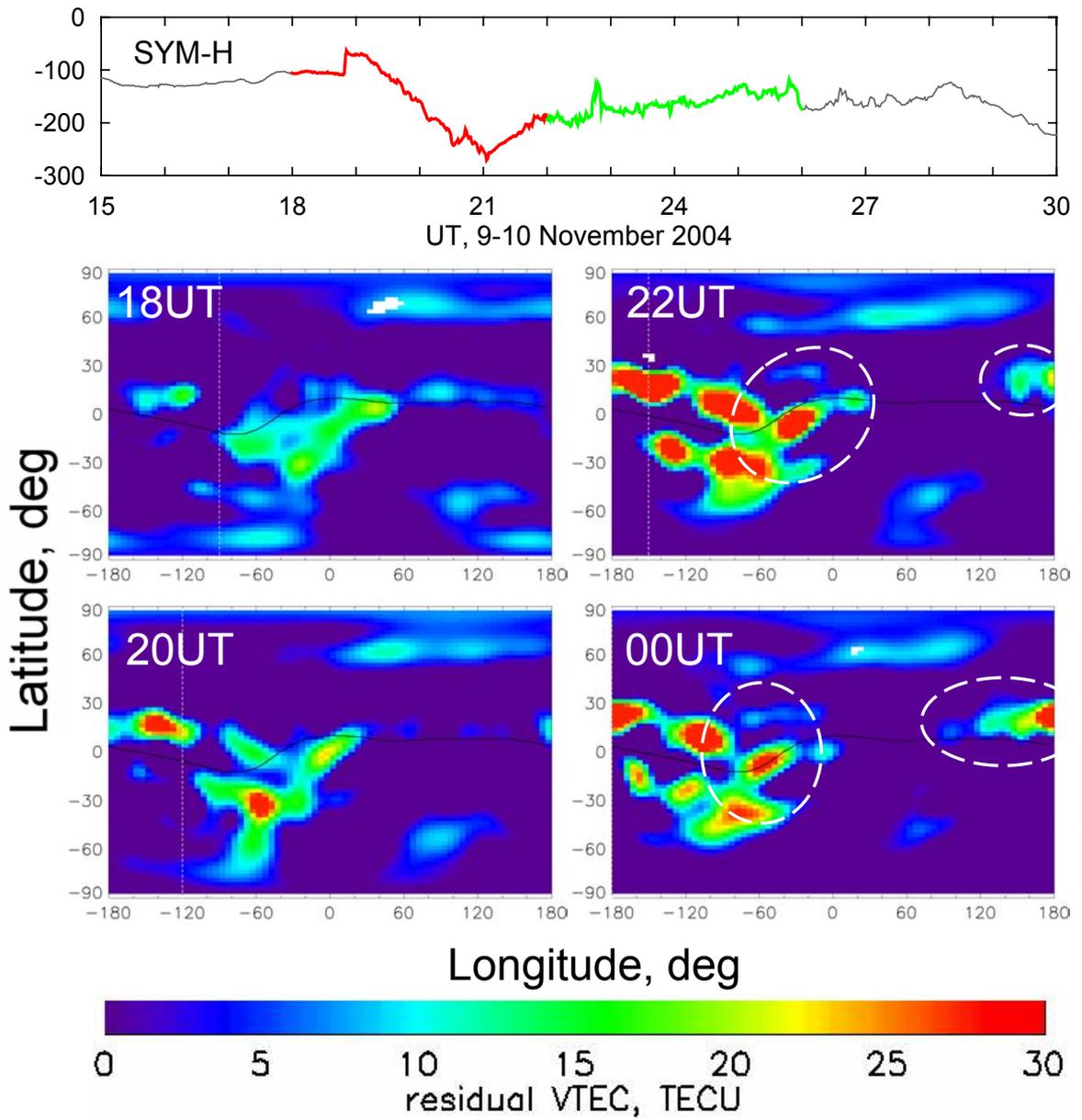

Figure 10. Same as Figure 6 but for 9-10 November 2004.

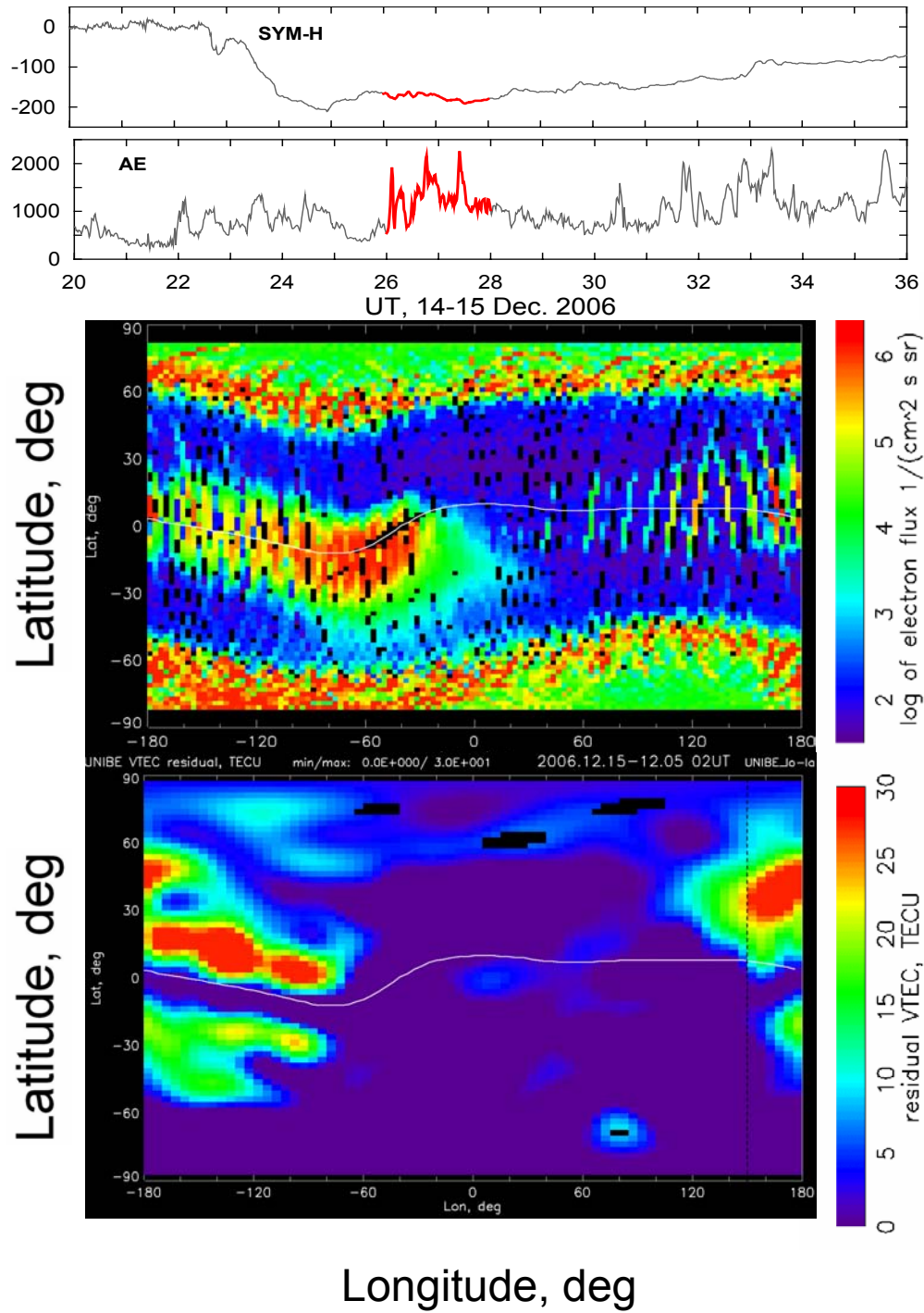

Figure 11. Great electron enhancements pattern (middle panel) and an example of positive ionospheric storm at 2-4 UT (bottom panel) of geomagnetic storm on 14-15 December 2006 (two upper panels with SYM-H and AE indices).

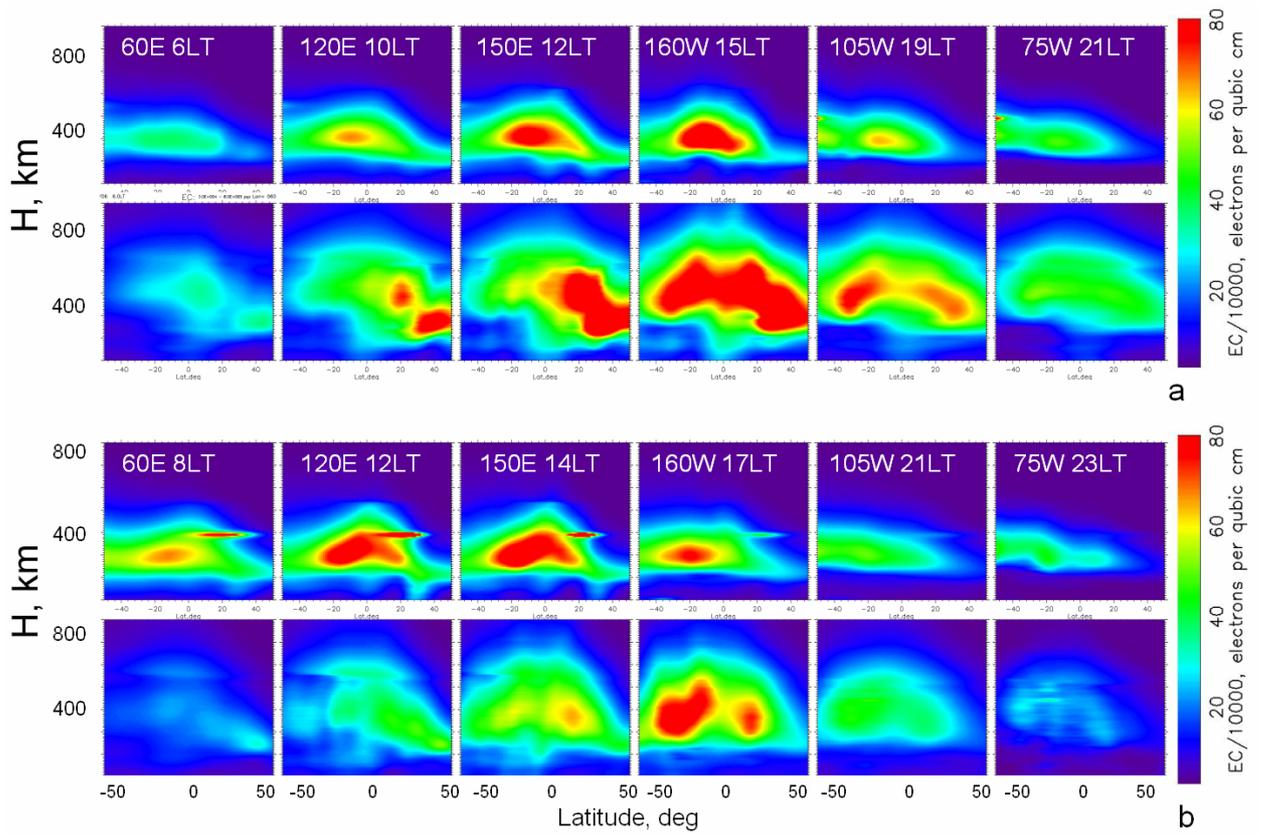

Figure 12. COSMIC/FS3 radio occultation tomography. Meridional cuts of EC for (a) 2 - 4 UT interval and (b) 4 - 6 UT for quiet day of 5 December 2006 (top) and storm-day of 15 December (bottom). Specific features are artifacts.

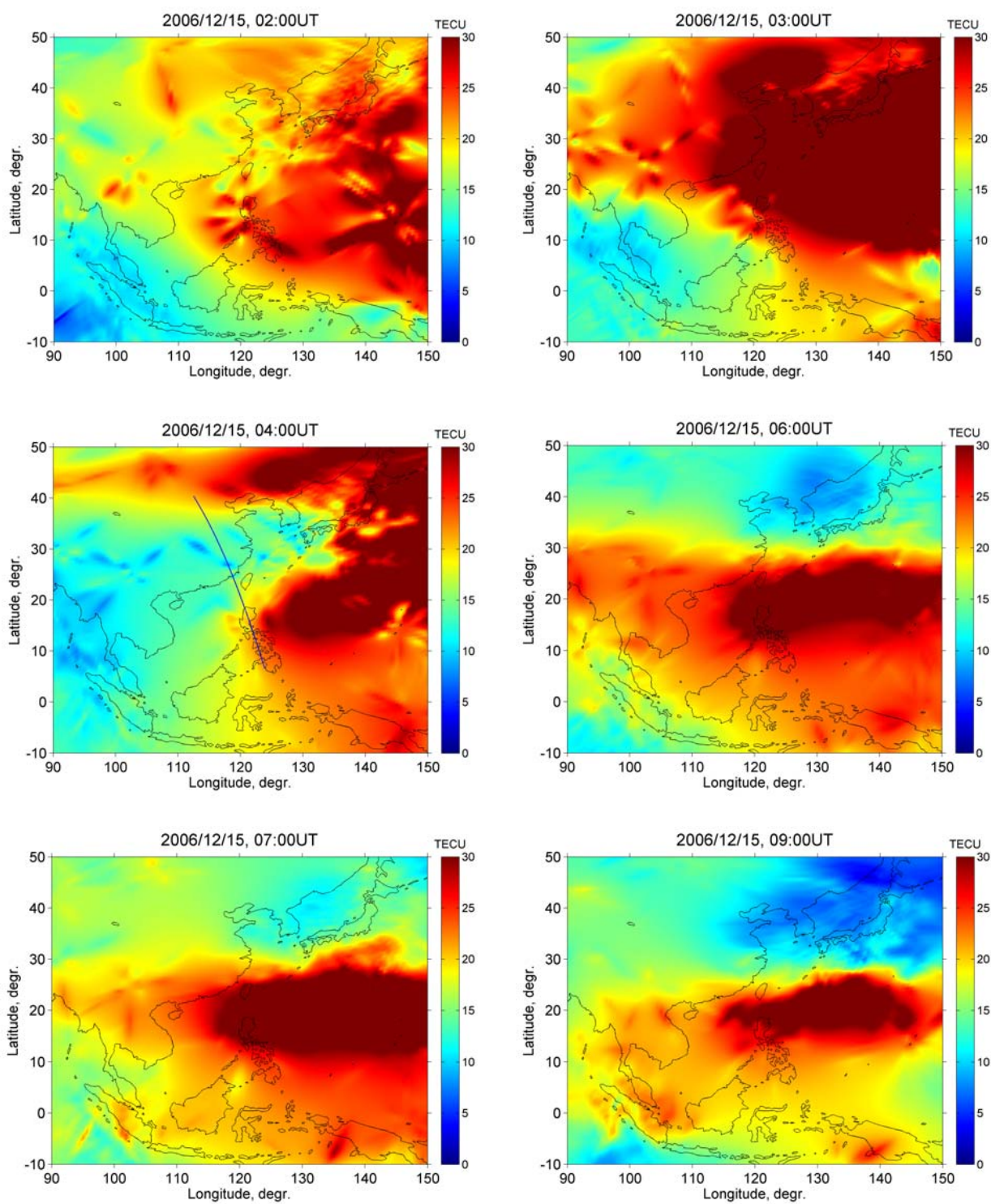

Figure 13. HORT reconstructions over Asian region on 15 December 2006 at 02:00 UT, 03:00 UT, 04:00UT, 06:00UT, 07:00UT, 09:00UT: vertical TEC maps in the latitude-longitude coordinates; the color scale is from 0 to 30 TECU. The blue line outlines the ground projection of FM1 satellite path (15 December 2006 at 03:59 UT).

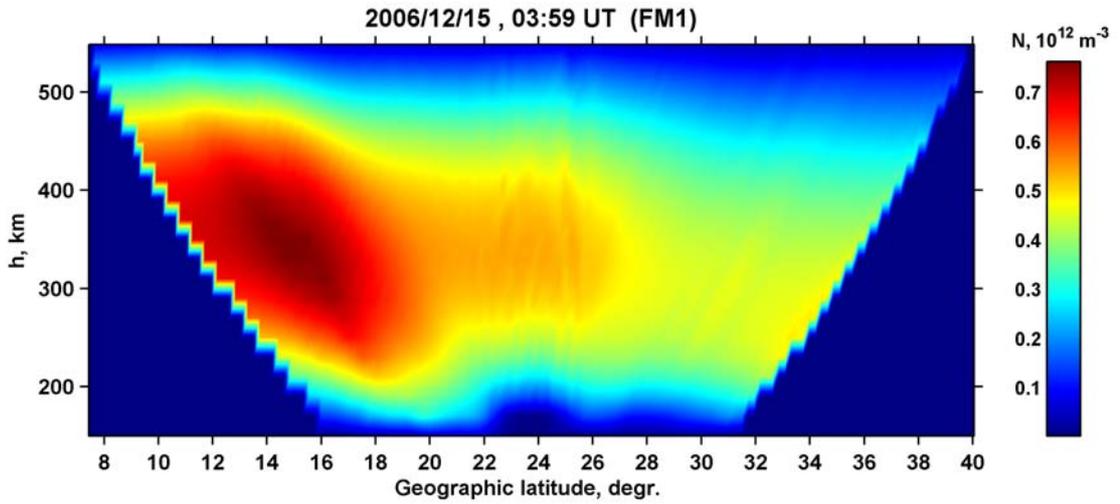

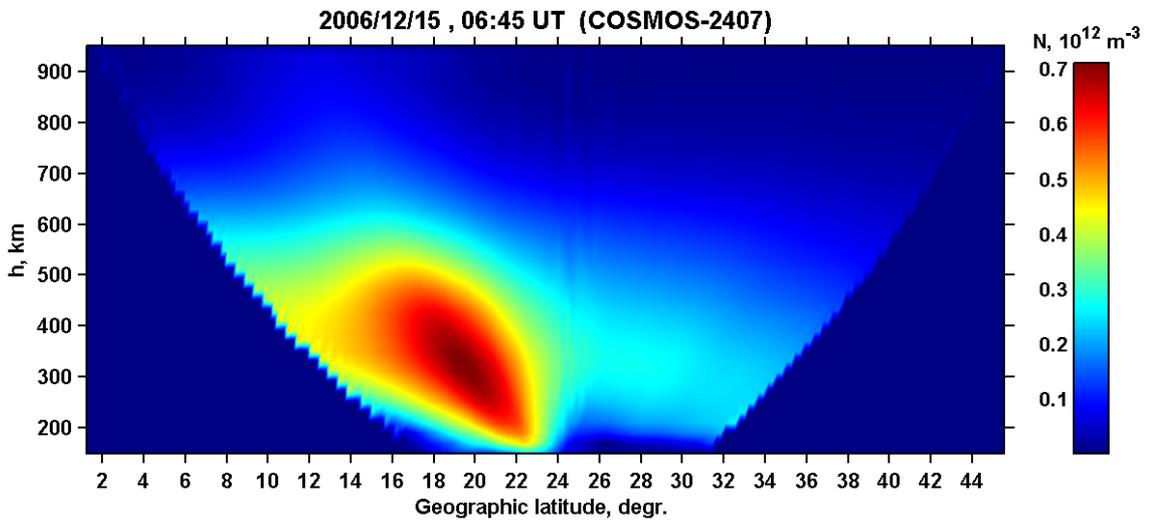

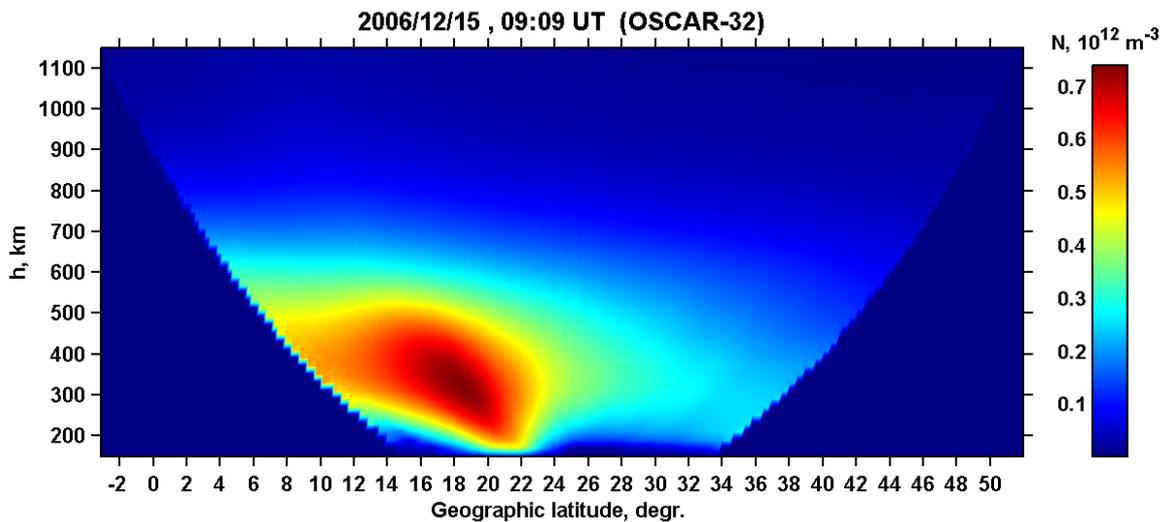

Figure 14. LORT image above Taiwan region as obtained from LITN data on 15 December 2006 at (a) 0359 UT, (b) 0645 UT and (c) 0909 UT.

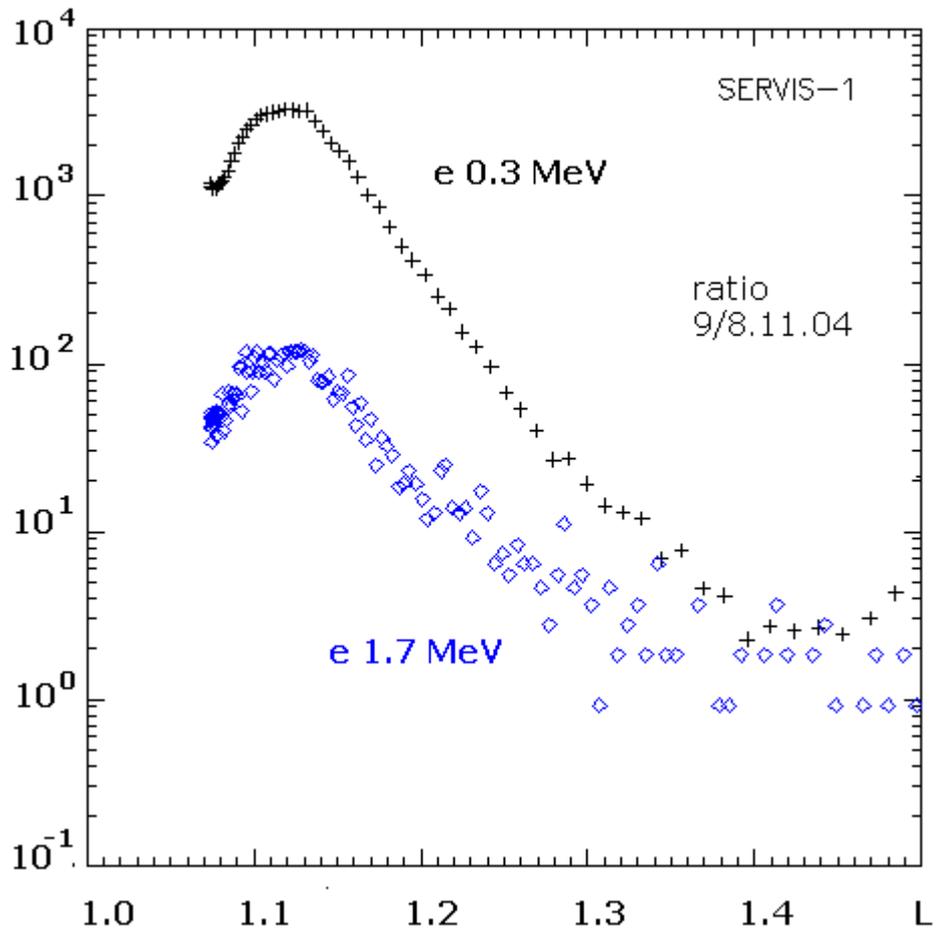

Figure 15. L-shell profiles of relative electron fluxes measured by SERVIS-1 in energy channels 0.3 - 1.5 and 1.7 - 3.4 MeV during storm-time period at ~21-22 UT on 9 November 2004. The fluxes are normalized to the quiet-day fluxes detected on 8 November 2004. A strong electron enhancement can be found at low L-shells during the magnetic storm.